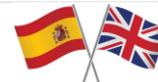

**MODELO HELIOS**

Eduardo Garbayo – Ingeniero Industrial
e-mail: yo@eduardogarbayo.com web: eduardogarbayo.com

*English translation on page 17*

# Hybrid Evaluation of Lifecycle and Impact of Outstanding Science
## Un Marco Predictivo Avanzado para la Madurez Tecnológica

**RESUMEN:** *Este paper presenta una mejora sustancial del modelo HELIOS (Hybrid Evaluation of Lifecycle and Impact of Outstanding Science), transformándolo de una herramienta estática de evaluación a un marco dinámico y predictivo para la madurez tecnológica. Se abordan las limitaciones del modelo original, que se basaba en normalización lineal y ponderaciones fijas. Las modificaciones clave incluyen la adopción de funciones de normalización no lineales (sigmoides), la integración de modelos de crecimiento en S para la previsión de indicadores clave (Inversión, Publicaciones, Patentes, Adopción, Regulación), la implementación de esquemas de ponderación dinámica basados en las fases del ciclo de vida, la aplicación de funciones de agregación no lineales para capturar sinergias y redundancias, y la incorporación de técnicas de cuantificación de la incertidumbre, como las simulaciones de Monte Carlo. Estas formulaciones matemáticas avanzadas permiten a HELIOS ofrecer pronósticos probabilísticos, identificar puntos de inflexión críticos y proporcionar una comprensión más matizada de la trayectoria tecnológica, lo que resulta invaluable para la planificación estratégica de la I+D, la evaluación de inversiones y la formulación de políticas en el ámbito de las tecnologías emergentes.*

**PALABRAS CLAVE**: Madurez tecnológica, Modelos predictivos, Curvas en S, Normalización no lineal, Ponderación dinámica, Integral de Choquet, Simulaciones de Monte Carlo, Incertidumbre.

**ABSTRACT.** *This paper presents a substantial enhancement of the HELIOS (Hybrid Evaluation of Lifecycle and Impact of Outstanding Science) model, transforming it from a static assessment tool into a dynamic and predictive framework for technological maturity. It addresses the limitations of the original model, which relied on linear normalization and fixed weights. Key modifications include the adoption of non-linear normalization functions (sigmoids), the integration of S-curve growth models for forecasting key indicators (Investment, Publications, Patents, Adoption, Regulation), the implementation of dynamic weighting schemes based on lifecycle phases, the application of non-linear aggregation functions to capture synergies and redundancies, and the incorporation of uncertainty quantification techniques such as Monte Carlo simulations. These advanced mathematical formulations enable HELIOS to provide probabilistic forecasts, identify critical inflection points, and offer a more nuanced understanding of a technology's trajectory. This makes it an invaluable tool for strategic planning, R&D investment evaluation, and policy-making in the domain of emerging technologies.*

## 1 Introducción

El modelo HELIOS original fue concebido como un marco integral para evaluar la madurez de tecnologías emergentes, integrando indicadores clave como la inversión en I+D (I), publicaciones científicas (P), patentes (Pt), nivel de adopción (A) y grado de regulación (R) [1]. Su objetivo era proporcionar un índice compuesto que reflejara el estado de desarrollo científico y tecnológico de una innovación. La formulación inicial del índice se basaba en una media ponderada de estas variables, normalizadas en el rango [0,1]:

$$\text{HELIOS} = w_I I_{\text{norm}} + w_P P_{\text{norm}} + w_{Pt} Pt_{\text{norm}} + w_A A_{\text{norm}} + w_R R_{\text{norm}}$$

donde $\sum w_i = 1$

Aunque útil para una evaluación puntual, el modelo original presentaba limitaciones significativas: una normalización lineal que no capturaba la dinámica no lineal de la madurez, ponderaciones estáticas que no se ajustaban a las fases del ciclo de vida, y una ausencia de capacidad predictiva cuantitativa explícita y cuantificación de la incertidumbre [1].

Este *paper* propone una serie de mejoras matemáticas para transformar HELIOS en un modelo predictivo robusto, capaz de ofrecer pronósticos probabilísticos y una comprensión más profunda de la trayectoria tecnológica.

## 2 Estado actual – Modelos predictivos conocidos

Para anticipar el recorrido de una tecnología emergente se han propuesto varios marcos conceptuales e indicadores cuantitativos. Un ejemplo muy difundido es el Ciclo de sobreexpectación (hype cycle) de Gartner, que describe cinco fases –«génesis», pico de sobreexpectación, valle de desilusión, pendiente de iluminación y meseta de productividad)– por las que supuestamente atraviesa cualquier innovación tecnológica. Sin embargo, este modelo es puramente cualitativo y basado en la percepción de mercado, por lo que carece de formulación matemática rigurosa. Sirve más para ilustrar tendencias generales de *hype* social que para realizar predicciones numéricas precisas. Aun así, en la práctica se utiliza para ubicar tecnologías según su visibilidad.

Otro enfoque es la curva en "S" de adopción/vida tecnológica, basada en modelos logísticos. Por ejemplo, la difusión de una tecnología suele representarse con una función logística acumulativa de ventas o usuarios. El modelo de difusión de Bass es un caso clásico: distingue innovadores (coeficiente $p$) e imitadores ($q$) y estima la adopción futura en función del mercado potencial. En dicho modelo los parámetros $p$ y $q$ (además del tamaño del mercado) determinan la forma de la curva de adopción. Estos modelos cuantitativos permiten ajustar datos históricos de ventas, usuarios o instalaciones, de modo que extrapolan cuándo la tecnología alcanzará su saturación o adopción masiva. La curva en "S" —que originalmente se usó para describir rendimiento o productividad frente al esfuerzo/duración– se adapta también a la adopción de producto (por ejemplo, categorías de innovadores, tempranos, mayoría temprana y tardía, etc., según Rogers). En la práctica, la "posicion" actual de la tecnología en la curva de adopción de Rogers (p.ej. si ya cruzó "el abismo" hacia la mayoría) y en la curva de Gartner se usan como indicadores de madurez.

El análisis bibliométrico y de patentes es otro método cuantitativo habitual. Aquí se rastrean las tendencias de publicaciones científicas, citas y registros de patentes asociadas a la tecnología. Por ejemplo, un fuerte crecimiento (posiblemente exponencial) en el número de artículos o patentes suele indicar que la disciplina está en expansión. La bibliometría permite detectar el "estado de la investigación" y sus tendencias, mientras que el conteo de patentes sirve para medir la actividad innovadora industrial. Ambos indicadores dominan las fuentes de datos en previsión tecnológica. Típicamente se puede ajustar una curva logística al crecimiento acumulado de publicaciones o patentes para estimar en qué año se alcanzaría un cierto nivel (p.ej. el 90 % del potencial). Complementariamente, se usan modelos econométricos o de series temporales (p.ej. ARIMA) sobre estos datos históricos para proyectar tendencias. En síntesis, las metodologías más robustas combinan varios indicadores medibles (I+D, publicaciones, patentes, adopción, etc.)

para extraer curvas de crecimiento y estimar la etapa futura de la tecnología.

No existe un único modelo cuantitativo universal; cada tecnología puede exigir distintas combinaciones. Por ejemplo, tecnologías científicas puras (física cuántica) pueden medirse mejor con publicaciones/patentes, mientras que tecnologías de consumo (IA, apps) se modelan mejor por adopción de mercado y factores sociales. En general se interpreta así: un incremento rápido en publicaciones y patentes anticipa crecimiento, mientras que un rápido crecimiento de adopción (ventas o usuarios) indica transición a etapa de expansión. La posición relativa entre estos indicadores permite ubicar la fase actual de la tecnología (investigación, piloto, nicho comercial, madurez, etc.).

## 3 Modelo HELIOS Elemental - Básico

Para cuantificar el nivel de madurez, definimos variables normalizadas $I, P, Pt, A, R$ (inversión, publicaciones, patentes, adopción, regulación) en el rango [0,1]. Cada variable se obtiene midiendo el dato empírico y aplicando una normalización adecuada (por ejemplo, dividiendo por un valor máximo histórico o meta). Entonces el índice HELIOS se computa como una **media ponderada** de estas variables:

$$\text{HELIOS} = w_I I + w_P P + w_{Pt} Pt + w_A A + w_R R,$$

donde los pesos $w_I, w_P, \ldots, w_R$ suman 1. Por ejemplo, se puede asignar aproximadamente $w_I = 0.25$, $w_P = 0.25$, $w_{Pt} = 0.20$, $w_A = 0.25$, $w_R = 0.05$, reflejando que inversión, publicaciones y adopción suelen ser determinantes, mientras que la regulación suele tener menor peso relativo en fases iniciales. El resultado HELIOS es un valor entre 0 (tecnología en fase muy incipiente) y 1 (madurez elevada). Este índice compuesto permite comparar tecnologías y evaluar su posición en el ciclo de vida: valores intermedios sugieren una fase de crecimiento, mientras valores cercanos a 1 indican cercanía a la saturación del mercado.

### 3.1 Criterios de puntuación y normalización

Para cada variable clave se definen criterios de medición y escalas estándar. A modo de ejemplo, se pueden plantear los siguientes parámetros:

**Inversión (I+D):** se mide el monto anual invertido (público + privado) en la tecnología (en USD). Se normaliza dividiendo por el mayor nivel registrado o meta sectorial (por ejemplo, un billón USD). Un rango típico de puntaje puede ser: 0 (casi 0 inversión), 0.2 (inversión baja), 0.5 (inversión moderada), 1.0 (inversión muy alta).

**Publicaciones científicas:** número de artículos académicos en la disciplina relacionada por año. Se normaliza tomando el valor observado vs. un máximo histórico. Por ejemplo, se puede establecer 0–1 según: 0–10 art (0–0.2), 11–50 (0.2–0.5), 51–200 (0.5–0.8), >200 (0.8–1.0). Un crecimiento exponencial de la bibliografía suele indicar fase de desarrollo rápido.

**Patentes:** número de familias de patentes publicadas anualmente en ese campo. Se normaliza similar a publicaciones. Criterios: 0–50 patentes (0–0.3), 51–200 (0.3–0.6), 201–500 (0.6–0.9), >500 (0.9–1.0). Por ejemplo, recientes análisis muestran miles de nuevas patentes de computación cuántica por año, lo que situaría a esa tecnología en puntajes altos para este indicador.

**Adopción:** grado de implementación o uso de la tecnología (por ejemplo, porcentaje de mercado o número de usuarios/pilotos). Se puede estimar como la penetración en el mercado objetivo. Un posible criterio: <1 % (score ~0), 1–10 % (0.1–0.3), 10–50 % (0.3–0.7), >50 % (0.7–1.0). La adopción real suele seguir una curva en $S$ (difusión de innovaciones), por lo que valores bajos implican fase inicial.

**Regulación:** nivel de madurez del marco legal/estándares. Se asigna en base cualitativa al porcentaje de aspectos regulados (0 = sin regulación; 0.5 = regulaciones parciales; 1.0 = regulación completa y armonizada). Por ejemplo, la presencia de guías o leyes específicas (como controles de exportación) puede valer 0.5–0.8, mientras la ausencia de regulación específica sería 0.0–0.2.

En resumen, cada indicador $X$ se normaliza a $x = \frac{(X - x_{min})}{(x_{max} - x_{min})}$ o mediante tramos definidos, obteniendo un valor de 0 a 1. Luego se pondera según la importancia asignada. La tabla siguiente ejemplifica criterios estándar (valores de referencia):

| Variable | Métrica | Normalización | Escala de puntaje aproximada |
|---|---|---|---|
| Inversión (USD) | Gasto anual en I+D | $I_{norm} = \frac{I}{I_{max}}$ | 0 (0) – 0.5 (~medio billón) – 1 (≥1 billón) |
| Publicaciones | Artículos/ano | $P_{norm} = \frac{P}{P_{max}}$ | 0 (0) – 0.5 (~100 pubs) – 1 (≥1000 pubs) |
| Patentes | Familias de patentes/ano | $Pt_{norm} = \frac{Pt}{Pt_{max}}$ | 0 (0) – 0.5 (~100 patentes) – 1 (≥500 patentes) |
| Adopción | % del mercado/usuarios | Escala fija | 0 (0%) – 0.2 (1–5%) – 0.5 (5–20%) – 1 (≥50%) |
| Regulación | Nivel regulatorio (0–1) | Directo (0 = nulo, 1 = completo) | 0 (nulo) – 0.5 (en desarrollo) – 1 (leyes/están dares) |

Estos rangos son orientativos y pueden ajustarse sectorialmente. Por ejemplo, la bibliometría y la estadística de patentes se usan como indicadores en "technology foresight" para identificar fases emergentes de tecnología. La justificación de normalizar frente a un máximo histórico o meta se basa en la idea de evaluar la fracción de progreso alcanzado.

### 3.2 Representación visual del modelo

El estado actual de las cinco variables se puede representar gráficamente. Por ejemplo, un **gráfico radar** mostrará cada variable en un eje radial (ver Figura abajo). Cada dimensión (inversión, publicaciones, patentes, adopción, regulación) está medida de 0 a 1, de modo que la superficie resultante refleja el perfil de madurez tecnológica. Asimismo, la típica **curva en S** ilustra la trayectoria de madurez global de la tecnología: su pendiente máxima al punto de inflexión (fase de rápido crecimiento) y el nivel de saturación final marca la madurez completa.

### 3.3 Ejemplo práctico: computación cuántica

Para ilustrar HELIOS, consideremos la **computación cuántica** con datos recientes. En términos de inversión, este sector ha atraído miles de millones de dólares; McKinsey estima que la innovación acelerada podría llevar el mercado cuántico global a unos \$100 mil millones en diez años. Supongamos que la inversión normalizada actual en cuántica es alta (por ejemplo $I = 0.8$). Respecto a *publicaciones*, el campo ha crecido exponencialmente; por ejemplo, en Estados Unidos hubo descensos en publicaciones generales de ciencia en 2023, pero la literatura cuántica ha aumentado, lo que podría corresponder a $P \approx 0.9$. Para *patentes*, los informes muestran miles de nuevas familias al año: por ejemplo, 3.795 en 2023, frente a 1.899 en 2020, indicando un campo con $Pt \approx 0.8$. La *adopción comercial* todavía es modesta (principalmente prototipos y servicios en nube), digamos $A = 0.3$. La *regulación* está emergiendo: en 2024 EE.UU. impuso controles de exportación específicos para tecnologías cuánticas, señal de atención normativa; podríamos asignar $R \approx 0.4$.

**Usando los pesos sugeridos, calculamos el índice HELIOS:**

$$HELIOS = 0.25 \cdot I + 0.25 \cdot P + 0.20 \cdot Pt + 0.25 \cdot A + 0.05 \cdot R$$
$$= 0.25(0.8) + 0.25(0.9) + 0.20(0.8) + 0.25(0.3) + 0.05(0.4) \approx 0.65.$$

Un valor de ~0.65 indica una etapa de crecimiento temprano, consistente con la rápida expansión de patentes e inversión, pero con adopción aún limitada. HELIOS en este caso resaltaría que la computación cuántica está aún lejos de la saturación; la pendiente creciente de la curva S sugeriría que el "tipping point" de adopción masiva podría estar por venir.

En conjunto, HELIOS ofrece un índice cuantitativo interpretativo: valores cercanos a 0.5–0.7 (como en este ejemplo) corresponderían a tecnologías en fase de desarrollo/adopción temprana, mientras que índices próximos a 1 significarían madurez o estancamiento (ya que la mayoría de indicadores se estabilizan o decrecen en madurez). Este ejemplo práctico demuestra cómo HELIOS permite integrar métricas empíricas recientes en un solo índice, facilitando comparaciones entre tecnologías y evaluaciones de su trayectoria futura.

# 4 Fundamentos Matemáticos para la Mejora del Modelo HELIOS

Para superar las limitaciones identificadas en el modelo HELIOS original, se propone la integración de metodologías matemáticas avanzadas que doten al marco de una mayor precisión, dinamismo y capacidad predictiva.

## 4.1 Normalización No Lineal de Indicadores

La normalización lineal empleada actualmente en HELIOS no captura adecuadamente la evolución no lineal de los indicadores de madurez tecnológica. La progresión de una tecnología, desde su concepción hasta su madurez, rara vez es lineal; más bien, sigue patrones de crecimiento que se aceleran y luego se desaceleran, formando una curva en S [2, 3]. Por lo tanto, se propone la adopción de **funciones sigmoides** para una representación más fiel de esta dinámica.

La normalización lineal asume que cada unidad de incremento en una métrica, como la inversión, tiene el mismo peso en la madurez, independientemente de si la tecnología se encuentra en una fase incipiente o madura [1]. Al utilizar funciones sigmoides, el modelo HELIOS puede reflejar que un pequeño aumento en la inversión o en las publicaciones en las primeras etapas, donde la curva de crecimiento es más plana, tiene un impacto diferente en la madurez que el mismo aumento durante la fase de crecimiento exponencial, donde la curva es más pronunciada, o cerca de la saturación, donde la curva vuelve a aplanarse [2, 4, 5, 6]. Esto proporciona una representación más matizada y empíricamente justificada del progreso de la madurez tecnológica.

Las funciones sigmoides, también conocidas como curvas en S, son modelos matemáticos que transforman cualquier valor de entrada real en un valor dentro de un rango acotado, típicamente entre 0 y 1 [4, 7]. Su forma de "S" es ideal para modelar procesos de crecimiento que inician lentamente, se aceleran rápidamente y luego se desaceleran hacia una saturación [2, 3, 4, 5, 6, 8]. Dos funciones sigmoides relevantes para esta aplicación son:

Función Logística Estándar: Definida como:

$$\sigma(x) = \frac{1}{1 + e^{-x}}$$

Para la normalización de un indicador $X_j$:

$$x_j = \frac{L_j}{1 + \exp(-k_j(X_j - X_{0j}))}$$

donde $L\_j$ es el límite superior (asíntota, típicamente 1), $k\_j$ es la tasa de crecimiento o pendiente de la curva para el indicador j, y $X\_0j$ es el valor del indicador j en el punto de inflexión [4, 8].

Función de Gompertz: Expresada como:

$$f(x) = a \cdot \exp(-b \cdot \exp(-c \cdot x))$$

Para la normalización de un indicador $X_j$:

$$x_j = a_j \cdot \exp(-b_j \cdot \exp(-c_j \cdot X_j))$$

donde a\_j es la asíntota superior (madurez máxima, típicamente 1), b\_j es el parámetro de desplazamiento a lo largo del eje x, y c\_j es la tasa de crecimiento [5, 9]. A diferencia de la logística, la función de Gompertz es una sigmoide asimétrica, caracterizada por un crecimiento más lento al inicio y al final del período, y una aproximación más gradual a la asíntota superior [5].

Esta asimetría la hace particularmente útil para modelar fenómenos donde el crecimiento inicial es muy lento antes de un despegue rápido, seguido de una saturación prolongada, como es común en la adopción de nuevas tecnologías o el crecimiento de poblaciones [5, 10].

Los parámetros específicos de estas funciones sigmoides $(L, k, X\_0 \text{ o } a, b, c)$ se estimarán empíricamente a partir de datos históricos de cada indicador mediante técnicas de regresión no lineal [9].

La elección entre la función logística (simétrica) y la de Gompertz (asimétrica) permite adaptar la normalización a las características específicas de cada indicador [5, 9]. Por ejemplo, la adopción de tecnologías a menudo sigue una curva de Gompertz, con una fase inicial muy lenta antes de un despegue rápido [5, 10], mientras que el crecimiento de publicaciones científicas podría ser más simétrico.

Esta flexibilidad en la elección de la función sigmoide para cada variable $(I, P, Pt, A, R)$ permite una calibración más precisa y una representación más fiel de su contribución individual a la madurez, mejorando la robustez general del modelo.

**Tabla 1: Comparación de Estrategias de Normalización de Indicadores para HELIOS**

| Indicador | Normalización Lineal Actual (Fórmula/Descripción) | Normalización Sigmoide Propuesta (Fórmula, Tipo) | Justificación de la Elección Sigmoide | Ejemplo (Valor Bruto) | Puntaje Normalizado (Lineal) | Puntaje Normalizado (Sigmoide) |
|---|---|---|---|---|---|---|
| Inversión | $I_{textnorm} = I/I_{textmax}$ (Lineal, tramos) [1] | $I\_textnorm = L/(1 + exp(-k(I - I\_0)))$ (Logística/Gompertz) | Captura crecimiento exponencial inicial y saturación de inversión [2, 11]. | $100M (de $1B máx) | 0.1 | 0.05 (si es fase temprana) |
| Publicaciones | $P\_textnorm = P/P\_textmax$ (Lineal, tramos) [1] | $P\_textnorm = L/(1 + exp(-k(P - P\_0)))$ (Logística/Gompertz) | Refleja crecimiento exponencial de la bibliografía en fases de desarrollo rápido [1, 12]. | 50 artículos (de 1000 máx) | 0.05 | 0.15 (si es fase acelerada) |
| Patentes | $Pt\_textnorm = Pt/Pt\_textmax$ (Lineal, tramos) [1] | $Pt\_textnorm = L/(1 + exp(-k(Pt - Pt\_0)))$ (Logística/Gompertz) | Modela la aparición gradual y posterior aceleración de patentes [1, 13, 14]. | 100 patentes (de 500 máx) | 0.2 | 0.3 (si es fase de crecimiento) |
| Adopción | Escala fija (0-1%)~0, (1-10%)~0.1-0.3, etc. [1] | $A\_textnorm = Mfrac1 - exp(-(p + q)t)1 + (q/p)exp(-(p + q)t)$ (Bass/Gompertz) | Refleja patrón de adopción asimétrico (curva en S) [1, 5, 10]. | 5% de mercado | 0.2-0.5 | 0.1 (si es fase inicial muy lenta) |
| Regulación | Directo (0=nulo, 0.5=desarrollo, 1=completo) [1] | $R\_textnorm = L/(1 + exp(-k(R - R\_0)))$ (Logística, con interpretación cualitativa) | Captura la progresión no lineal de la madurez regulatoria [1]. | Guías específicas (0.5-0.8) | 0.5-0.8 | 0.6 (si es fase intermedia) |

## 4.2 Modelos de Crecimiento y Difusión (Curvas en S)

Para dotar a HELIOS de una capacidad predictiva robusta, es esencial modelar las trayectorias de crecimiento de sus indicadores clave utilizando funciones en S, que son inherentes a la evolución tecnológica [2, 3, 15]. Estas curvas se caracterizan por una fase inicial lenta (génesis), una fase de crecimiento rápido (aceleración) y una fase final de desaceleración y saturación (madurez) [2, 3, 8].

Las curvas en S no solo describen el crecimiento acumulado, sino que sus derivadas proporcionan información crucial sobre la *velocidad* y *aceleración* del desarrollo tecnológico [11, 13, 16]. El **punto de inflexión** de la curva en S, donde la primera derivada es máxima, representa el "punto de despegue" o "tipping point" de la tecnología, indicando la fase de mayor crecimiento [1, 8, 17]. La pendiente de la curva en S también puede revelar la "productividad de las inversiones en I+D" [11]. Al cuantificar estos puntos y tasas de cambio, HELIOS se transforma de una herramienta descriptiva a una predictiva con hitos temporales claros, lo que es invaluable para la toma de decisiones estratégicas de inversión y desarrollo.

La modelación de cada indicador clave (Publicaciones, Patentes, Adopción) con su propia curva en S permite una visión granular y temporal de la madurez tecnológica [1]. Esto es superior a un único índice compuesto, ya que revela desequilibrios. Por ejemplo, una tecnología podría mostrar un pico en publicaciones y patentes, pero una adopción aún incipiente. Esta situación indicaría que la tecnología se encuentra en una fase de "valle de la desilusión" o "desarrollo temprano" a pesar del avance científico [1]. Esta capacidad de diagnóstico multifacético proporciona una comprensión más profunda de la trayectoria de la tecnología y ayuda a identificar áreas que requieren mayor atención o inversión para superar "el abismo" hacia la adopción masiva [1, 18].

Para la **Adopción (A)**, el **Modelo de Difusión de Bass** es un enfoque clásico y ampliamente utilizado para pronosticar la adopción de nuevos productos e innovaciones [1, 10, 19, 20]. Este modelo distingue entre dos tipos de adoptantes: innovadores (p) e imitadores (q) [19, 20]. La tasa de adopción se formula como:

Para la **Adopción (A)**, el **Modelo de Difusión de Bass**:

$$\frac{dN(t)}{dt} = \left[p + q\frac{N(t)}{M}\right][M - N(t)]$$

donde N(t) es el número acumulado de adoptantes en el tiempo t, y M es el tamaño total del mercado potencial o nivel de saturación [10, 19, 20]. La adopción acumulada se expresa como:

La adopción acumulada se expresa como:

$$N(t) = M\frac{1 - \exp(-(p+q)t)}{1 + (q/p)\exp(-(p+q)t)}$$

El punto de inflexión, que indica el momento de máxima tasa de adopción, se calcula como $t\_textinflexion = fracln(q/p)p + q$ [19]. Los parámetros p, q y M se estiman a partir de datos históricos de adopción mediante regresión no lineal [10, 19, 20].

Para las **Publicaciones Científicas (P)** y las **Patentes (Pt)**, cuyo crecimiento también sigue una curva en S [1, 12, 13, 14, 21], se pueden emplear **Modelos Logísticos o de Gompertz**. La formulación del Modelo Logístico para Y(t) (número acumulado de publicaciones o patentes) es:

Para las **Publicaciones (P)** y **Patentes (Pt)**, Modelo Logístico:

$$Y(t) = \frac{K}{1 + \exp(-r(t - t_0))}$$

donde K es la capacidad de carga o nivel de saturación, r es la tasa de crecimiento y $t\_0$ es el tiempo en el punto de inflexión [8, 12, 21]. Para el Modelo de Gompertz, la formulación es:

Modelo de Gompertz:

$$Y(t) = a \cdot \exp(-b \cdot \exp(-c \cdot t))$$

con a como asíntota superior, b como parámetro de desplazamiento y c como tasa de crecimiento [5, 9]. Los parámetros de estos modelos se estiman utilizando regresión no lineal sobre datos históricos [12, 14, 21].

Una vez que los parámetros de los modelos en S se han estimado a partir de datos históricos, estas funciones pueden extrapolarse para pronosticar los valores futuros de I, P, Pt y A. Esto permite generar proyecciones cuantitativas para cada indicador, que a su vez alimentarán un índice HELIOS proyectado a lo largo del tiempo.

**Tabla 2: Parámetros y Pronósticos Clave de los Modelos de Curva en S para Indicadores HELIOS**

| Indicador | Modelo de Curva en S Elegido | Parámetros Clave Estimados (Ejemplo) | Rango de Datos Históricos Utilizados | Año Pronosticado del Punto de Inflexión | Nivel de Saturación Pronosticado (Valor y Año Estimado) | Valor Pronosticado a 5 Años |
|---|---|---|---|---|---|---|
| Publicaciones | Logístico | K=1500, r=0.2, t0=2028 | 2000-2024 | 2028 | 1500 artículos (2040) | 1000 artículos |
| Patentes | Gompertz | a=700, b=15, c=0.1 | 2005-2024 | 2030 | 700 familias (2045) | 450 familias |
| Adopción | Bass | M=100%, p=0.01, q=0.3 | 2010-2024 | 2032 | 100% de mercado (2050) | 25% de mercado |
| Inversión | Logístico | K=$50B, r=0.15, t0=2027 | 2015-2024 | 2027 | $50 mil millones (2042) | $35 mil millones |

**Figura 1: Curvas de Pronóstico de Publicaciones, Patentes y Adopción**

*Descripción Conceptual:* Esta figura mostraría tres gráficos de línea superpuestos o adyacentes, cada uno representando la trayectoria pronosticada de un indicador (Publicaciones, Patentes, Adopción) a lo largo del tiempo, utilizando sus respectivos modelos de curva en S. Cada curva debería resaltar su punto de inflexión y la asíntota de saturación.

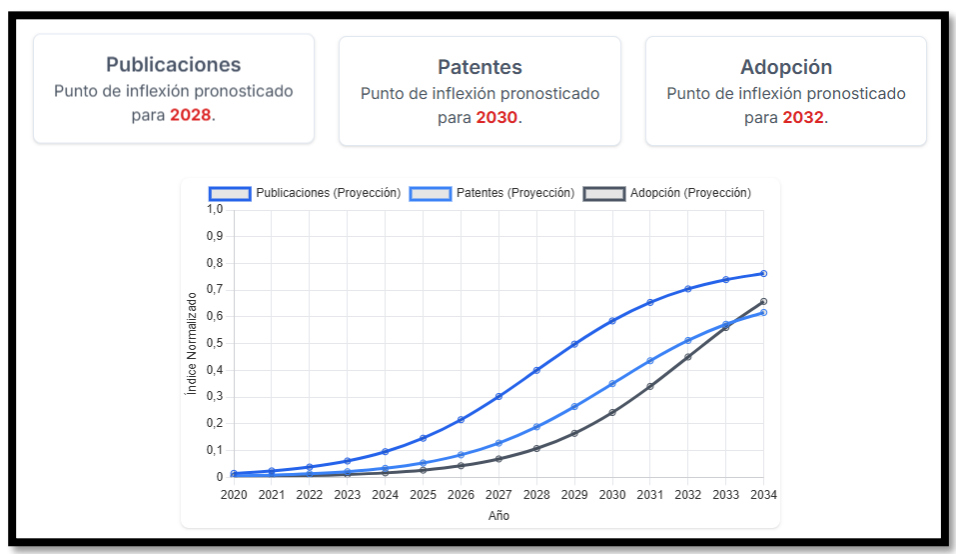

*Ilustración 1Pronóstico con Curvas de Crecimiento -Utilizando modelos de difusión como el de Bass y funciones logísticas, HELIOS proyecta la trayectoria futura de cada indicador*

### 4.3 Ponderación Dinámica y Basada en Datos

La asignación de pesos estáticos en HELIOS es una simplificación que no refleja la importancia cambiante de los indicadores a lo largo del ciclo de vida de una tecnología. Los pesos fijos asignados en el modelo HELIOS original no se ajustan a la evolución de la tecnología [1]. La relevancia de la inversión o las publicaciones es muy alta en las fases iniciales, mientras que la adopción y la regulación ganan peso a medida que la tecnología madura y se difunde [1, 22]. Se propone, por tanto, un esquema de **ponderación dinámico y basado en datos**.

La adopción de pesos dinámicos y basados en datos representa un salto cualitativo para HELIOS. En lugar de una evaluación estática, el modelo se vuelve *sensible al contexto* de la tecnología [23, 24, 25]. Esto significa que la importancia de, por ejemplo, una nueva patente, se evaluará de manera diferente si la tecnología está en su fase de investigación, donde una patente es un hito enorme, que si está en una fase de madurez, donde las patentes pueden ser incrementales. Este enfoque adaptativo no solo mejora la precisión de la evaluación de la madurez, sino que también permite una comprensión más profunda de los impulsores clave en cada etapa del ciclo de vida tecnológico, lo cual es fundamental para la toma de decisiones estratégicas.

Para determinar los pesos, se pueden emplear **métodos de ponderación basados en datos** como el Análisis de Componentes Principales (PCA) o el Análisis Factorial (FA). PCA es una técnica de reducción de dimensionalidad que identifica patrones subyacentes en conjuntos de variables correlacionadas y puede derivar pesos objetivos basándose en la proporción de varianza explicada por los componentes principales [26, 27, 28]. Las "cargas" (loadings) de cada factor en el primer componente principal (PC1) pueden representar los pesos, calculados como:

Para PCA:

$$w_j = \frac{|loading_{j,PC1}|}{\sum_{k=1}^{n}|loading_{k,PC1}|}$$

donde n es el número de indicadores [27]. FA, similar a PCA, se enfoca en la identificación de variables latentes que influyen en las variables observadas y también se utiliza para derivar pesos en índices compuestos [28, 29]. La elección entre PCA y Análisis Factorial para la derivación de pesos no es meramente técnica, sino que tiene implicaciones teóricas [28]. PCA busca maximizar la varianza explicada y reducir la dimensionalidad. Si la "madurez tecnológica" se conceptualiza como una construcción latente que impulsa los indicadores, FA podría ofrecer una validación teórica más sólida para los pesos. Si el objetivo principal es la reducción de datos y la identificación de las combinaciones de indicadores que mejor resumen la información, PCA podría ser más directo. Un análisis experto consideraría esta distinción para asegurar que el método de ponderación elegido se alinee con la conceptualización teórica del modelo HELIOS.

Además, se implementarán **esquemas de ponderación adaptativos (dinámicos)**, donde los pesos se ajustan en función de la fase del ciclo de vida de la tecnología [23, 24, 25, 30]. Esto permite que el modelo refleje con mayor precisión la importancia cambiante de cada indicador a medida que la tecnología madura [1]. Las fases del ciclo de vida (por ejemplo, "Emergencia/Concepto", "Desarrollo Inicial", "Adopción Temprana", "Crecimiento/Expansión", "Madurez/Saturación") se definirán basándose en umbrales del índice HELIOS o de indicadores individuales [1, 31, 32].

Para cada fase, se aplicará un conjunto diferente de pesos ($w\_text{fase,i}$). Por ejemplo, en la fase de emergencia, se asignaría un mayor peso a la inversión y las publicaciones, y un menor peso a la adopción y la regulación. En la fase de crecimiento, la adopción y las patentes ganarían importancia, mientras que la regulación aumentaría su peso.

En la madurez, la adopción y la regulación serían más determinantes, y la inversión, publicaciones y patentes podrían estabilizarse o decrecer [1]. Estos pesos por fase pueden determinarse mediante juicio experto refinado, donde los expertos asignan pesos específicos para *cada fase* del ciclo de vida en lugar de un único conjunto estático [24, 25, 29]. Si se dispone de suficientes datos históricos de diversas tecnologías a lo largo de sus ciclos de vida, se pueden emplear algoritmos de aprendizaje automático o técnicas de optimización para aprender los pesos óptimos para cada fase, lo que podría implicar modelos predictivos que ajustan los pesos en tiempo real [33, 34].

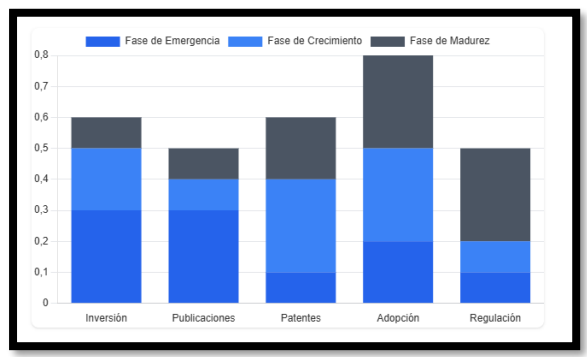

*Ilustración 2 Ponderación Dinámica y Adaptativa - La importancia de cada indicador cambia a lo largo del ciclo de vida de una tecnología. HELIOS ajusta los pesos dinámicamente para reflejar qué factores son más críticos en cada fase.*

**Tabla 3: Esquema de Ponderación Dinámica Propuesto para HELIOS por Fase del Ciclo de Vida**

| Fase del Ciclo de Vida Tecnológica | Pesos Propuestos para Inversión (I) | Pesos Propuestos para Publicaciones (P) | Pesos Propuestos para Patentes (Pt) | Pesos Propuestos para Adopción (A) | Pesos Propuestos para Regulación (R) | Justificación Racional |
|---|---|---|---|---|---|---|
| **Emergencia/ Concepto** [1] | 0.35 | 0.30 | 0.10 | 0.05 | 0.05 | Inversión y publicaciones son críticas para la viabilidad inicial y la exploración del conocimiento [1]. |
| **Desarrollo Inicial** [1] | 0.30 | 0.30 | 0.15 | 0.10 | 0.05 | Énfasis en la investigación y el desarrollo temprano, con creciente actividad de patentamiento [1]. |
| **Adopción Temprana** [1] | 0.20 | 0.20 | 0.25 | 0.30 | 0.05 | La adopción comienza a ser un factor clave, junto con la protección de la propiedad intelectual [1]. |
| **Crecimiento /Expansión** [1] | 0.15 | 0.15 | 0.20 | 0.40 | 0.10 | La adopción masiva impulsa el crecimiento; la regulación empieza a ser relevante para la escalabilidad [1]. |
| **Madurez/Sat uración** [1] | 0.10 | 0.10 | 0.15 | 0.35 | 0.30 | El mercado se estabiliza, la regulación y los estándares son fundamentales para la consolidación [1]. |

*Nota: Los pesos son conceptuales y deben ser calibrados con datos empíricos y/o juicio experto refinado.*

**Figura 2: Ponderación Dinámica de Indicadores por Fase del Ciclo de Vida**

*Descripción Conceptual:* Esta figura ilustraría cómo los pesos de cada indicador (Inversión, Publicaciones, Patentes, Adopción, Regulación) cambian a través de las diferentes fases del ciclo de vida tecnológico (Emergencia, Crecimiento, Madurez). Podría ser un gráfico de barras apiladas o un gráfico de radar para cada fase, mostrando la distribución de pesos.

# 5 Funciones de Agregación No Lineales y Modelado de Sinergias

La agregación lineal simple en HELIOS, que utiliza un promedio ponderado ($HELIOS = \sum w_i x_i$), asume una relación compensatoria perfecta entre los indicadores [1, 35, 36]. Esto implica que un bajo rendimiento en un indicador puede ser completamente compensado por un alto rendimiento en otro. Sin embargo, en la madurez tecnológica, esta suposición no siempre es válida; por ejemplo, la ausencia de un marco regulatorio esencial no puede ser "compensada" por una inversión masiva si la tecnología requiere un entorno legal para su adopción generalizada. Además, la agregación lineal no modela sinergias, donde el efecto combinado de varios indicadores es mayor que la suma de sus partes [36, 37]. Para un modelado más sofisticado, se explorarán operadores de agregación no lineales. La madurez tecnológica no es simplemente una suma lineal de sus componentes; existen interdependencias complejas, efectos de umbral y fenómenos de sinergia o redundancia [35, 36]. Al pasar de una media ponderada lineal a operadores de agregación no lineales como la Integral de Choquet o los operadores OWA, HELIOS puede reflejar con mayor precisión estas complejidades [36, 38, 39]. Esto permite al modelo distinguir, por ejemplo, entre una tecnología que avanza de manera equilibrada en todos los frentes, lo cual podría ser valorado más positivamente por la Integral de Choquet, y una que tiene un rendimiento excepcional en un área, pero deficiencias críticas en otra. Esta capacidad de modelar relaciones no compensatorias es fundamental para una evaluación de madurez verdaderamente experta y matizada.

Entre los **operadores de agregación no lineales avanzados**, se destacan:

**Integral de Choquet:** Este es un operador de agregación no aditivo que permite modelar interacciones (sinergias o redundancias) entre los criterios [36, 38]. Utiliza "medidas difusas" o "capacidades" que asignan importancia no solo a criterios individuales, sino también a sus combinaciones [38]. Una ventaja clave es que puede penalizar logros desequilibrados, por ejemplo, una alta inversión pero una baja adopción, y capturar sinergias. Por ejemplo, una alta inversión y un marco regulatorio emergente podrían tener un efecto super-aditivo en la madurez percibida, ya que la regulación reduce el riesgo de la inversión [36]. La fórmula conceptual para un conjunto finito de indicadores es:

$$C_\mu(x_1, \ldots, x_n) = \sum_{i=1}^{n} x_{(i)} \left[ \mu(A_{(i)}) - \mu(A_{(i+1)}) \right]$$

donde $x_{(i)}$ son los valores normalizados ordenados de forma no decreciente, $A_{(i)}$ es el conjunto de los $i$ indicadores más grandes, y $\mu$ es una medida difusa que asigna un valor a cada subconjunto de indicadores [38].

**Operadores de Promedio Ponderado Ordenado (OWA - Ordered Weighted Averaging):** Esta es una clase parametrizada de operadores de agregación que permite modelar instrucciones de agregación lingüísticas, como "al menos la mitad de los criterios se cumplen" [39, 40]. Los operadores OWA ofrecen un espectro de comportamientos de agregación, desde el mínimo hasta el máximo, incluyendo la media aritmética y la mediana, ajustando un conjunto de pesos [39]. Permiten incorporar una "actitud" (optimista o pesimista) en la agregación [39]. La fórmula es:

$$OWA(a_1, \ldots, a_n) = \sum_{j=1}^{n} w_j \, b_j$$

donde $b_j$ es el j-ésimo valor más grande de los argumentos $a_i$ (los valores normalizados de los indicadores), y $w_j$ son los pesos OWA que suman 1 [39].

**Agregación Basada en Lógica Difusa:** Este enfoque permite manejar información imprecisa o cualitativa, como la del indicador de regulación [41, 42, 43]. Utiliza reglas de inferencia (por ejemplo, SI (Inversión es ALTA Y Adopción es BAJA) ENTONCES Madurez es CRECIMIENTO TEMPRANO) para combinar los valores de los indicadores [42]. Sus ventajas incluyen la robustez para datos inciertos e imprecisos, la capacidad de modelar relaciones complejas y no lineales sin necesidad de funciones matemáticas explícitas, y su interpretabilidad [42, 43]. La aplicación de la lógica difusa o de medidas difusas dentro de la Integral de Choquet podría transformar la entrada cualitativa de "Regulación" en una representación más matizada y matemáticamente tratable [41, 42, 43]. En lugar de un simple 0, 0.5 o 1, se podrían asignar grados de pertenencia a conjuntos difusos como "Regulación Incipiente", "Regulación en Desarrollo" o "Regulación Consolidada". Esto no solo reduce la subjetividad, sino que también permite que la información sobre la regulación interactúe de manera más sofisticada con los demás indicadores en el proceso de agregación, mejorando la coherencia matemática del modelo.

**Tabla 4: Comparación de Métodos de Agregación para el Índice HELIOS**

| Método de Agregación | Fórmula Matemática (Conceptual) | Propiedades Clave/Ventajas | Desventajas/Complejidad | Relevancia para HELIOS |
|---|---|---|---|---|
| **Media Aritmética Ponderada** | $textHELIOS = sum w_i x_i$ [1] | Simple, fácil de calcular e interpretar. | Asume compensación perfecta, no modela sinergias/redundancias [35, 36]. | Base actual, útil para comparaciones rápidas si las interacciones son mínimas. |
| **Integral de Choquet** | $C\_mu(x\_1, dots, x\_n)$ $sum x\_(i)[mu(A\_(i)) - mu(A\_(i+1))]$ [38] | Maneja sinergias y redundancias, penaliza desequilibrios, permite medidas difusas [36, 38]. | Requiere estimación de medidas difusas (capacidades), más compleja [38]. | Ideal para capturar interdependencias complejas entre I, P, Pt, A, R y evaluar madurez equilibrada. |
| **Operadores OWA** | $textOWA(a\_1, dots, a$ $sum w\_j b\_j$ [39] | Permite modelar actitudes (optimista/pesimista), flexible entre min y max [39]. | Requiere ordenamiento de valores, selección de pesos OWA [39]. | Útil para incorporar preferencias en la agregación o reflejar diferentes estrategias de evaluación. |
| **Lógica Difusa** | Basado en reglas "SI-ENTONCES" y funciones de pertenencia [42, 43] | Robusta para datos inciertos/cualitativos, interpretable, modela relaciones no lineales implícitamente [42, 43]. | Requiere definición de reglas y conjuntos difusos, puede ser intensiva en datos [42]. | Mejora la cuantificación de la Regulación y permite una agregación más intuitiva de la información cualitativa. |

## 6 Pronóstico Probabilístico y Cuantificación de la Incertidumbre

Para que el modelo HELIOS sea verdaderamente predictivo y útil para la toma de decisiones estratégicas, debe incorporar la cuantificación de la incertidumbre, proporcionando pronósticos probabilísticos en lugar de estimaciones puntuales. Los pronósticos de punto único no reflejan la incertidumbre inherente en las predicciones, especialmente en campos complejos como la previsión tecnológica [44, 45]. La Cuantificación de la Incertidumbre (UQ) es esencial para comprender el rango de posibles resultados, la confianza en las predicciones y para identificar qué incertidumbres son más críticas [44, 45].

La integración de pronósticos probabilísticos y la cuantificación de la incertidumbre eleva el modelo HELIOS de una herramienta de evaluación a un sistema de apoyo a la decisión de alto nivel [44, 45]. Los tomadores de decisiones no solo sabrán *cuál* es el nivel de madurez proyectado, sino también *cuán seguros* pueden estar de esa predicción y *cuál es el rango de posibles trayectorias futuras*. Este enfoque puede resultar especialmente útil para la gestión de riesgos en contextos de alta incertidumbre, la asignación de recursos y la formulación de estrategias en entornos de alta incertidumbre tecnológica, permitiendo decisiones más informadas y resilientes.

Entre los **métodos para la Cuantificación de la Incertidumbre en HELIOS** se incluyen:

**Simulaciones de Monte Carlo (MCS):** MCS es un enfoque basado en muestreo ampliamente utilizado para la cuantificación y propagación de incertidumbres [44, 46, 47, 48, 49]. Permite generar una distribución de posibles valores del índice HELIOS futuro, en lugar de un único valor [44, 46]. El proceso implica:

**Definición de Distribuciones de Incertidumbre:** Se definirán distribuciones de probabilidad para los parámetros de entrada del modelo (por ejemplo, incertidumbre en $p, q, M$ del modelo de Bass; o en $K, r, t\_0$ de los modelos logísticos; o en los valores normalizados de los indicadores basados en la varianza histórica o en la elicitación de expertos) [46, 47].

**Muestreo Repetido:** Se realizarán miles o millones de simulaciones, extrayendo valores aleatorios de estas distribuciones de incertidumbre para cada parámetro [46, 47].

**Cálculo del Índice HELIOS:** Para cada simulación, se calculará el índice HELIOS completo, utilizando la normalización no lineal, los pesos dinámicos y la agregación no lineal.

**Distribución de Resultados**: La colección resultante de valores HELIOS formará una distribución de probabilidad que representa el pronóstico probabilístico del índice [46]. Las ventajas de MCS incluyen su facilidad de interpretación, robustez y capacidad para manejar distribuciones no gaussianas y entradas correlacionadas [44, 46, 47]. Los resultados permiten generar intervalos de predicción (por ejemplo, un intervalo de confianza del 95% para el índice HELIOS) y estimar la probabilidad de alcanzar ciertos umbrales de madurez [45, 48, 50].

**Métodos Bayesianos:** Estos métodos proporcionan un marco riguroso para la estimación de parámetros y la UQ, incorporando conocimiento previo (priors) y actualizando las creencias con nuevos datos [51, 52]. Pueden estimar directamente las distribuciones posteriores de los parámetros del modelo (por ejemplo, p,q del modelo de Bass), que luego se utilizan para generar pronósticos probabilísticos. Su ventaja radica en ofrecer una UQ más completa, especialmente útil cuando los datos son limitados, al permitir la incorporación de información experta o de análogos a través de las distribuciones previas [47, 51].

El **Análisis de Sensibilidad** es una extensión de la UQ que identifica qué incertidumbres de entrada (por ejemplo, la variabilidad en un peso específico o la incertidumbre en un parámetro de una curva en S) son las que más contribuyen a la incertidumbre en la salida del índice HELIOS [48, 49]. Esto ayuda a enfocar los esfuerzos de mejora del modelo y la recopilación de datos en las fuentes de incertidumbre más significativas [44, 49]. A menudo se realiza simultáneamente con la UQ utilizando métodos de Monte Carlo [49].

La literatura sobre pronósticos sugiere que combinar predicciones de diferentes métodos o modelos, como diferentes modelos de curvas en S para el mismo indicador o diferentes enfoques de ponderación, puede mejorar significativamente la precisión y reducir los errores [53, 54]. Esta estrategia de "pronóstico combinado" o "ensemble forecasting" [53] puede aplicarse a HELIOS: en lugar de elegir un único modelo para cada componente (normalización, S-curvas, ponderación, agregación), se podrían ejecutar múltiples variaciones y combinar sus resultados probabilísticos. Esto añade una capa adicional de sofisticación y robustez, mitigando el riesgo de depender de un único conjunto de suposiciones o la calibración de un modelo específico.

**Figura 3: Pronóstico Probabilístico del Índice HELIOS a lo Largo del Tiempo**

*Descripción Conceptual:* El eje X representa el tiempo en años, mientras que el eje Y representa el Índice HELIOS, que varía de 0 a 1. La línea central sólida muestra la mediana del pronóstico del índice HELIOS, indicando la trayectoria más probable de la madurez tecnológica. Alrededor de esta línea central, se visualizan áreas sombreadas que representan intervalos de predicción, como el 50% y el 95% de confianza. Estas áreas sombreadas ilustran el rango de resultados posibles para el índice HELIOS en cada momento futuro, reflejando la incertidumbre inherente en las predicciones. La amplitud de estas áreas sombreadas puede variar, ensanchándose a medida que el pronóstico se extiende más en el futuro, lo que indica un aumento de la incertidumbre. Esta visualización es crucial para comunicar la incertidumbre de manera efectiva a los tomadores de decisiones, ya que proporciona una comprensión más completa de los riesgos y oportunidades asociados con la trayectoria de una tecnología.

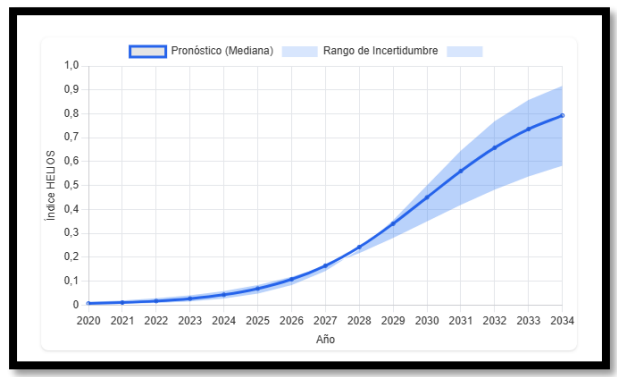

*Ilustración 3 Agregación Avanzada y Cuantificación de Incertidumbre El modelo utiliza la Integral de Choquet para modelar sinergias y penalizar desequilibrios. Mediante simulaciones de Monte Carlo, HELIOS proporciona un rango de futuros posibles, permitiendo una mejo*

# 7 Propuesta de un Modelo HELIOS Mejorado: Formulación Matemática Integral

El modelo HELIOS mejorado integra las formulaciones matemáticas avanzadas propuestas para ofrecer una evaluación, pronóstico y cuantificación de la incertidumbre más precisos de la madurez tecnológica. El índice HELIOS (H) en un momento t se define como una función de agregación no lineal de los indicadores normalizados dinámicamente, con pesos adaptativos:

$$H(t) = \textbf{Aggregation\_Function}(x_I(t), x_P(t), x_{Pt}(t), x_A(t), x_R(t) | W_{\textbf{dynamic}}(t))$$

Donde textAggregation_Function puede ser la Integral de Choquet, un operador OWA o una función de agregación basada en lógica difusa, seleccionada en función de las propiedades de interacción deseadas entre los indicadores. W_textdynamic(t) representa el conjunto de pesos adaptativos para la fase de madurez de la tecnología en el tiempo t.

El algoritmo completo para el cálculo, pronóstico y cuantificación de la incertidumbre del índice HELIOS mejorado se estructura en los siguientes pasos:

1. **Normalización No Lineal de Indicadores**:
   Para cada indicador X_j (Inversión, Publicaciones, Patentes, Adopción, Regulación), se aplica una función sigmoide f_j (Logística o Gompertz), calibrada con datos históricos. Por ejemplo, para una normalización logística:

   $$x_j(t) = \frac{L_j}{1 + \exp(-k_j(X_j(t) - X_{0j}))}$$

   Los parámetros $L\_j, k\_j, X\_0j$ se estiman previamente para cada indicador j a partir de sus datos históricos.

2. **Modelado de Crecimiento y Pronóstico de Indicadores (Curvas en S)**:
   Para los indicadores de Publicaciones (P), Patentes (Pt) y Adopción (A), se ajustan modelos de curva en S (Logístico, Gompertz o Bass) a sus datos históricos. Estos modelos permiten pronosticar los valores futuros de $X\_j(t)$ para ttexttiempo_actual.

   **Adopción (A): Se utiliza el Modelo de Bass:**

   $$A(t) = M_A \frac{1 - \exp(-(p_A + q_A)t)}{1 + (q_A/p_A)\exp(-(p_A + q_A)t)}$$

   **Publicaciones (P): Se puede emplear un Modelo Logístico:**

   $$P(t) = \frac{K_P}{1 + \exp(-r_P(t - t_{0P}))}$$

   **Patentes (Pt): Un Modelo Logístico o Gompertz también es adecuado:**

   $$P(t) = \frac{K_{Pt}}{1 + \exp(-r_{Pt}(t - t_{0Pt}))}$$

   Los parámetros $(M\_A, p\_A, q\_A, K\_P, r\_P, t\_0P, K\_Pt, r\_Pt, t\_0Pt)$ se calibran con datos históricos.

3. **Determinación de Pesos Dinámicos**:
   Se identifica la fase actual de la tecnología (e.g., "Emergencia", "Crecimiento", "Madurez") basándose en el valor actual del índice HELIOS o en umbrales de indicadores clave. Se aplica el conjunto de pesos predefinido o derivado por PCA/FA para esa fase.

   $$Wdynamic(t) = \{wI, fase, wP, fase, wPt, fase, wA, fase, wR, fase\}$$

4. **Agregación No Lineal**:
   Los valores normalizados $x\_j(t)$ y los pesos dinámicos $W\_textdynamic(t)$ se combinan utilizando una función de agregación no lineal. Por ejemplo, si se elige la Integral de Choquet, se requiere la definición de una medida difusa mu que capture las interacciones entre los indicadores.

5. **Pronóstico Probabilístico y Cuantificación de la Incertidumbre:**
   Para generar pronósticos probabilísticos y cuantificar la incertidumbre, se emplean Simulaciones de Monte Carlo.

   - Se definen distribuciones de probabilidad para los parámetros de los modelos de curva en S ($e.g., p\_A, q\_A, M\_A, etc.$), los parámetros de normalización ($L\_j, k\_j, X\_0j$), y, si aplica, las medidas difusas o los pesos OWA.
   - Se realizan N iteraciones ($e.g.$, 100,000) de la simulación. En cada iteración k:
     - Se muestrean aleatoriamente los parámetros de las distribuciones definidas.
     - Se pronostican los valores de los indicadores $X\_j(t)\_k$ para el horizonte de tiempo deseado.
     - Se normalizan $x\_j(t)\_k$.
     - Se determina la fase de la tecnología y se aplican los pesos dinámicos $W\_textdynamic(t)\_k$.
     - Se calcula el índice $H(t)\_k$ utilizando la función de agregación no lineal.
   - Los N valores resultantes de $H(t)\_k$ para cada t forman una distribución de probabilidad, a partir de la cual se pueden derivar la mediana, los intervalos de confianza (e.g., 95%) y otras métricas de incertidumbre.

6. **Análisis de Sensibilidad:**

   Posteriormente, se realiza un análisis de sensibilidad sobre los resultados de Monte Carlo para identificar cuáles de las incertidumbres en los parámetros de entrada tienen mayor impacto en la variabilidad del índice HELIOS pronosticado.

# 8 Conclusiones y recomendaciones

La evolución del modelo HELIOS, desde su formulación original hasta el marco avanzado propuesto, representa una evolución metodológica relevante respecto al modelo anterior en la evaluación y previsión de la madurez tecnológica. Al superar las limitaciones de la linealidad, la estaticidad y la falta de capacidad predictiva, el modelo se convierte en una herramienta más precisa para la toma de decisiones estratégicas. La normalización no lineal mediante funciones sigmoides refleja con mayor fidelidad la contribución real de cada indicador, dado que el impacto de un cambio en una métrica varía según la fase del ciclo de vida: un avance inicial o un incremento en saturación tienen significados distintos al mismo cambio en fase de crecimiento exponencial.

La integración de modelos de crecimiento en S para publicaciones, patentes y adopción aporta capacidad predictiva cuantitativa, permitiendo pronosticar puntos de inflexión y niveles de saturación, fundamentales para planificar I+D y anticipar "puntos de despegue". Esta visión granular también diagnostica desequilibrios entre indicadores y áreas críticas para superar barreras de adopción.

La ponderación dinámica, ajustada a cada fase del ciclo de vida, mejora la relevancia de la evaluación y optimiza la asignación de recursos. Finalmente, las funciones de agregación no lineales y la cuantificación de incertidumbre mediante simulaciones de Monte Carlo capturan interacciones y sinergias, ofreciendo no solo predicciones, sino rangos probabilísticos con niveles de confianza, esenciales para la gestión de riesgos en entornos de alta incertidumbre.

## 8.1 Recomendaciones:

**Recopilación y Calibración de Datos Históricos:** Podría ser beneficioso establecer un programa robusto para la recopilación continua de datos históricos detallados para cada indicador (I, P, Pt, A, R) a lo largo del tiempo y para diversas tecnologías. Estos datos son la base para la calibración precisa de los parámetros de las funciones sigmoides, los modelos de crecimiento en S y las medidas difusas.

**Desarrollo de Software Específico:** Se recomienda el desarrollo de una plataforma de software dedicada que implemente las formulaciones matemáticas avanzadas propuestas. Esta plataforma debería permitir la entrada de datos, la calibración de modelos, la ejecución de simulaciones de Monte Carlo, la visualización de pronósticos probabilísticos y la realización de análisis de sensibilidad.

**Validación y Refinamiento Continuo:** El modelo debe ser sometido a un proceso de validación y refinamiento continuo. Esto incluye la comparación de los pronósticos del modelo con la evolución real de las tecnologías, la revisión periódica de los pesos dinámicos y las funciones de agregación, y la incorporación de nuevos datos y conocimientos expertos para mejorar la precisión y la robustez.

**Elicitación de Expertos para Parámetros y Medidas Difusas:** Para maximizar la precisión, especialmente en ausencia de datos históricos suficientes, se recomienda la elicitación estructurada de expertos para definir las distribuciones de incertidumbre de los parámetros iniciales de los modelos y para construir las medidas difusas en la Integral de Choquet, si se opta por esta.

**Análisis de Escenarios y Sensibilidad:** Utilizar la capacidad de cuantificación de la incertidumbre para realizar análisis de escenarios. Esto permitirá a los tomadores de decisiones explorar el impacto de diferentes suposiciones (ej. cambios en la inversión, nuevas regulaciones) en la trayectoria de madurez de una tecnología y comprender qué factores son los más críticos para su éxito o fracaso.

Al implementar estas mejoras, el modelo HELIOS se posiciona como una herramienta interesante y con posibilidad de mejora para la previsión y gestión de la madurez tecnológica, proporcionando una base cuantitativa y probabilística para decisiones estratégicas en el dinámico panorama de la innovación.

## 9 Modelo abierto

Las propuestas de mejora que hemos esbozado representan una hoja de ruta para llevar el modelo HELIOS a su siguiente fase de evolución. Partiendo de la base matemáticamente rigurosa ya establecida, estas ampliaciones buscan transformar el modelo de un marco estático a una herramienta predictiva, dinámica y robusta, capaz de reflejar la complejidad inherente al ciclo de vida de la ciencia.

Para alcanzar este objetivo, consideramos fundamental una colaboración continua y abierta. En primer lugar, la validación y calibración rigurosa del modelo es un paso crítico. Proponemos el uso de técnicas de backtesting y validación cruzada para asegurar que el modelo no solo funcione en teoría, sino que también tenga una capacidad predictiva demostrable con datos históricos. Además, la incorporación de un enfoque bayesiano para la calibración de sus parámetros nos permitirá no solo hacer predicciones, sino también cuantificar la incertidumbre asociada a ellas, proporcionando una base más sólida para la toma de decisiones.

En segundo lugar, la riqueza de la información reside no solo en los números, sino también en el contexto. Por ello, sugerimos la integración de datos no estructurados y análisis de redes. El uso de técnicas de Procesamiento del Lenguaje Natural (PNL) para analizar textos de publicaciones, patentes y noticias podría extraer nuevas variables latentes, como el sentimiento o la dirección de la investigación. Al mismo tiempo, el análisis de redes, como las de coautoría y citaciones, revelaría la estructura y el dinamismo de la comunidad científica, identificando a los actores clave y las interconexiones que impulsan la innovación.

Finalmente, para superar las limitaciones de las formulaciones fijas, proponemos la adopción de modelos de aprendizaje automático. En lugar de depender de una única fórmula, modelos de ensamble como Random Forest o Gradient Boosting pueden capturar las complejas interacciones no lineales entre los indicadores.

Además, para una gestión óptima de carteras de inversión en ciencia, la optimización multiobjetivo nos permitiría ir más allá de una simple puntuación, equilibrando múltiples factores como el impacto potencial, el riesgo y la diversidad tecnológica.

Estamos convencidos de que estas ampliaciones, en colaboración con la comunidad, pueden transformar HELIOS en una herramienta avanzada. Por ello, invitamos a la comunidad académica, a los expertos en análisis de datos y a los profesionales del sector a unirse a nosotros para validar, refinar y expandir este modelo. Juntos, podemos construir un sistema de evaluación que no solo mida el pasado, sino que también proporciona un marco mejorado para el análisis prospectivo en ciencia y tecnología.

## 10 Declaración sobre el uso de iA

Durante la elaboración de este documento se empleó un modelo de lenguaje de código abierto (LLMO), ejecutado en un entorno local, correspondiente a la versión disponible en mayo de 2025. Esta herramienta se utilizó para ampliar y refinar el modelo HELIOS básico, generando borradores y explorando formulaciones alternativas. No intervino en el análisis crítico ni en la interpretación de resultados. Asimismo, se realizaron pruebas con datos reales utilizando este mismo modelo, con el objetivo de obtener resultados experimentales que se incorporaron también en la publicación del libro del mismo autor: Horizontes Disruptivos: Las 10 tecnologías que redefinirán la próxima década (ISBN 9798294856489). Todo el contenido generado fue revisado, editado por el autor.

## 11 REFERENCIAS

# HELIOS:

**H**ybrid **E**valuation of **L**ifecycle and **I**mpact of **O**utstanding **S**cience
Un Marco Predictivo Avanzado para la Madurez Tecnológica.

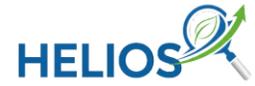

| The English version of this document is a translation of the Spanish original | 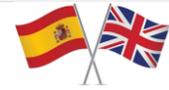 English Version |
| --- | --- |

# HELIOS MODEL

Eduardo Garbayo – Industrial Engineer
e-mail: yo@eduardogarbayo.com
web: eduardogarbayo.com

*Hybrid Evaluation of Lifecycle and Impact of Outstanding Science*
*An Advanced Predictive Framework for Technology Maturity*

**RESUMEN:** *This paper presents a substantial improvement of the HELIOS (Hybrid Evaluation of Lifecycle and Impact of Outstanding Science) model, transforming it from a static evaluation tool to a dynamic and predictive framework for technology maturity. It addresses the limitations of the original model, which was based on linear normalization and fixed weights. Key modifications include the adoption of nonlinear normalization functions (sigmoids), the integration of S-growth models for forecasting key indicators (Investment, Publications, Patents, Adoption, Regulation), the implementation of dynamic weighting schemes based on life cycle phases, the application of nonlinear aggregation functions to capture synergies and redundancies, and the incorporation of uncertainty quantification techniques such as Monte Carlo simulations. These advanced mathematical formulations enable HELIOS to provide probabilistic forecasts, identify critical inflection points and provide a more nuanced understanding of the technology trajectory, which is invaluable for strategic R&D planning, investment assessment and policy formulation in the field of emerging technologies..*

**KEYWORDS**: Technological maturity, Predictive models, S-curves, Non-linear normalization, Dynamic weighting, Choquet Integral, Monte Carlo simulations, Uncertainty, etc..

**ABSTRACT.** *This paper presents a substantial enhancement of the HELIOS (Hybrid Evaluation of Lifecycle and Impact of Outstanding Science) model, transforming it from a static assessment tool into a dynamic and predictive framework for technological maturity. It addresses the limitations of the original model, which relied on linear normalization and fixed weights. Key modifications include the adoption of non-linear normalization functions (sigmoids), the integration of S-curve growth models for forecasting key indicators (Investment, Publications, Patents, Adoption, Regulation), the implementation of dynamic weighting schemes based on lifecycle phases, the application of non-linear aggregation functions to capture synergies and redundancies, and the incorporation of uncertainty quantification techniques such as Monte Carlo simulations. These advanced mathematical formulations enable HELIOS to provide probabilistic forecasts, identify critical inflection points, and offer a more nuanced understanding of a technology's trajectory. This makes it an invaluable tool for strategic planning, R&D investment evaluation, and policy-making in the domain of emerging technologies.*

**KEYWORDS**: *Technological maturity, Predictive models, S-curves, Non-linear normalization, Dynamic weighting, Choquet integral, Monte Carlo simulations, Uncertainty quantification..*

## 12 Introduction

The original HELIOS model was conceived as a comprehensive framework for assessing the maturity of emerging technologies, integrating key indicators such as R&D investment (I), scientific publications (P), patents (Pt), level of adoption (A) and degree of regulation (R) [1]. Its objective was to provide a composite index reflecting the state of scientific and technological development of an innovation. The initial formulation of the index was based on a weighted average of these variables, normalized to the range [0.1]:

$$\text{HELIOS} = w_I I_{\text{norm}} + w_P P_{\text{norm}} + w_{Pt} Pt_{\text{norm}} + w_A A_{\text{norm}} + w_R R_{\text{norm}}$$

where $\sum w_i = 1$

Although useful for a point-in-time assessment, the original model had significant limitations: a linear normalization that did not capture the nonlinear dynamics of maturity, static weights that did not match life-cycle phases, and an absence of explicit quantitative predictive capability and uncertainty quantification [1].

This paper proposes a series of mathematical improvements to transform HELIOS into a robust predictive model, capable of providing probabilistic forecasts and a deeper understanding of the technological trajectory..

# 13 Current status - Known predictive models

Several conceptual frameworks and quantitative indicators have been proposed to anticipate the path of an emerging technology. One widely used example is Gartner's hype cycle, which describes five phases - "genesis, peak of hype, valley of disillusionment, slope of enlightenment and plateau of productivity" - through which any technological innovation supposedly passes. However, this model is purely qualitative and based on market perception, so it lacks rigorous mathematical formulation. It serves more to illustrate general social hype trends than to make precise numerical predictions. Even so, in practice it is used to rank technologies according to their visibility.

Another approach is the technology adoption/lifetime "S" curve, based on logistic models. For example, the diffusion of a technology is usually represented by a cumulative logistic function of sales or users. The Bass diffusion model is a classic case: it distinguishes innovators (coefficient p) and imitators (q) and estimates future adoption as a function of the potential market. In such a model, the parameters p and q (in addition to market size) determine the shape of the adoption curve. These quantitative models allow fitting historical data on sales, users or installations, so that they extrapolate when the technology will reach saturation or mass adoption. The "S" curve - originally used to describe performance or productivity versus effort/duration - is also adapted to product adoption (e.g., innovator, early, early majority, late majority, etc., according to Rogers). In practice, the current "position" of the technology on the Rogers adoption curve (e.g., whether it has already crossed "the chasm" to majority) and on the Gartner curve are used as indicators of maturity.

Bibliometric and patent analysis is another common quantitative method. Here, trends in scientific publications, citations and patent filings associated with the technology are tracked. For example, strong (possibly exponential) growth in the number of articles or patents usually indicates that the discipline is expanding. Bibliometrics allows the detection of the "state of research" and its trends, while patent counts serve to measure industrial innovative activity. Both indicators dominate the data sources in technology forecasting. Typically, a logistic curve can be fitted to the cumulative growth of publications or patents to estimate in which year a certain level (e.g. 90% of the potential) would be reached. Complementarily, econometric or time series models (e.g. ARIMA) are used on these historical data to project trends. In summary, the most robust methodologies combine several measurable indicators (R&D, publications, patents, adoption, etc.) to extract growth curves and estimate the future stage of the technology.

There is no single universal quantitative model; each technology may require different combinations. For example, pure scientific technologies (quantum physics) may be best measured by publications/patents, while consumer technologies (AI, apps) are best modeled by market adoption and social factors. In general it is interpreted as follows: a rapid increase in publications and patents anticipates growth, while a rapid growth in adoption (sales or users) indicates transition to expansion stage. The relative position between these indicators allows to locate the current phase of the technology (research, pilot, commercial niche, maturity, etc.).

# 14 HELIOS Elemental Model – Basic

To quantify the level of maturity, we define normalized variables I,P,Pt,A,R (investment, publications, patents, adoption, regulation) in the range [0,1]. Each variable is obtained by measuring the empirical data and applying an appropriate normalization (e.g., dividing by a historical maximum or target value). The HELIOS index is then computed as a weighted average of these variables:

$$\text{HELIOS} = w_I I + w_P P + w_{Pt} Pt + w_A A + w_R R,$$

where the weights w_I,w_P,...,w_R add up to 1. For example, one can assign approximately w_I=0.25, w_P=0.25, w_Pt=0.20, w_A=0.25, w_R=0.05 , reflecting that investment, publications and adoption tend to be determinant, while regulation tends to have less relative weight in early stages. The HELIOS result is a value between 0 (very early stage technology) and 1 (high maturity). This composite index makes it possible to compare technologies and evaluate their position in the life cycle: intermediate values suggest a growth phase, while values close to 1 indicate near market saturation.

## 14.1 Scoring and standardization criterio

Standard measurement criteria and scales are defined for each key variable. By way of example, the following parameters can be considered:

**Investment (I+D):** measures the annual amount invested (public + private) in the technology (in USD). It is normalized by dividing by the highest recorded level or sectoral target (e.g. one billion USD). A typical range of scores can be: 0 (almost 0 investment), 0.2 (low investment), 0.5 (moderate investment), 1.0 (very high investment).

**Scientific publications:** number of academic articles in the related discipline per year. It is normalized by taking the observed value vs. a historical maximum. For example, 0-1 can be set according to: 0-10 art (0-0.2), 11-50 (0.2-0.5), 51-200 (0.5-0.8), >200 (0.8-1.0). Exponential growth of the literature usually indicates a phase of rapid development.

**Patents:** number of patent families published annually in that field. It is normalized similar to publications. Criteria: 0-50 patents (0-0.3), 51-200 (0.3-0.6), 201-500 (0.6-0.9), >500 (0.9-1.0). For example, recent analyses show thousands of new quantum computing patents per year,

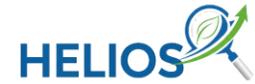

which would place that technology in high scores for this indicator.

**Adoption:** degree of implementation or use of the technology (e.g., percentage of market or number of users/pilots). It can be estimated as the penetration in the target market. One possible criterion: <1 % (score ~0), 1-10 % (0.1-0.3), 10-50 % (0.3-0.7), >50 % (0.7-1.0). Actual adoption usually follows an S-curve (diffusion of innovations), so low values imply initial phase.

**Regulation:** level of maturity of the legal framework/standards. It is assigned on a qualitative basis to the percentage of regulated aspects (0 = no regulation; 0.5 = partial regulations; 1.0 = complete and harmonized regulation). For example, the presence of specific guidelines or laws (such as export controls) may be worth 0.5-0.8, while the absence of specific regulation would be 0.0-0.2..

In summary, each indicator X is normalized to $x = \frac{(X - x_{\min})}{(x_{\max} - x_{\min})}$ or by means of defined sections, obtaining a value from 0 to 1, and then weighted according to the importance assigned. The following table exemplifies standard criteria (reference values):

| Variable | Metrica | Standardization | Approximate scoring scale |
|---|---|---|---|
| Investment (USD) | Annual expenditure on I+D | $I_{\text{norm}} = \frac{I}{I_{\max}}$ | 0 (0) – 0.5 (~half billón) – 1 (≥1 billón) |
| Publications | Articles/year | $P_{\text{norm}} = \frac{P}{P_{\max}}$ | 0 (0) – 0.5 (~100 pubs) – 1 (≥1000 pubs) |
| Patents | Patent families/year | $Pt_{\text{norm}} = \frac{Pt}{Pt_{\max}}$ | 0 (0) – 0.5 (~100 patentes) – 1 (≥500 patents) |
| Adoption | % of the market/users | Fixed scale | 0 (0%) – 0.2 (1–5%) – 0.5 (5–20%) – 1 (≥50%) |
| Regulation | Regulatory level (0–1) | Direct (0 = void, 1 = complete) | 0 (void) – 0.5 (in development) – 1 (laws/standards) |

These ranges are indicative and can be adjusted sectorally. For example, bibliometrics and patent statistics are used as indicators in technology foresight to identify emerging phases of technology. The rationale for normalizing against a historical maximum or target is based on the idea of assessing the fraction of progress achieved.

## 14.2 Visual representation of the model

The current state of the five variables can be plotted graphically. For example, a radar chart will show each variable on a radial axis (see Figure below). Each dimension (investment, publications, patents, adoption, regulation) is measured from 0 to 1, so that the resulting surface reflects the technology maturity profile. Likewise, the typical S-curve illustrates the overall technology maturity trajectory: its maximum slope indicates the inflection point (rapid growth phase) and the final saturation level marks full maturity.

## 14.3 Practical Example: Quantum Computing

To illustrate HELIOS, let's consider quantum computing with recent data. In terms of investment, this sector has attracted billions of dollars; McKinsey estimates that accelerated innovation could drive the global quantum market to about $100 billion in ten years. Assume that current normalized investment in quantum is high (e.g. I=0.8 ). With respect to publications, the field has grown exponentially; for example, in the United States there were declines in general science publications in 2023, but the quantum literature has increased, which could correspond to P≈0.9 . For patents, reports show thousands of new families per year: for example, 3,795 in 2023[, versus 1,899 in 2020, indicating a field with Pt≈0.8 . Commercial adoption is still modest (mainly prototypes and cloud services), say A=0.3 . Regulation is emerging: in 2024 the US imposed specific export controls for quantum technologies, a sign of regulatory attention; we could assign R≈0.4.

**Using the weights suggested, we calculate the HELIOS index:**

HELIOS $= 0.25 \cdot I + 0.25 \cdot P + 0.20 \cdot Pt + 0.25 \cdot A + 0.05 \cdot R$
$\square$ $= 0.25(0.8) + 0.25(0.9) + 0.20(0.8) + 0.25(0.3) + 0.05(0.4) \approx 0.65.$

A value of ~0.65 indicates an early growth stage, consistent with rapid expansion of patents and investment, but with still limited adoption. HELIOS in this case would highlight that quantum computing is still far from saturation; the increasing slope of the S-curve would suggest that the "tipping point" of mass adoption may be ahead.

Overall, HELIOS provides an interpretive quantitative index: values close to 0.5-0.7 (as in this example) would correspond to technologies in the development/early adoption phase, while indices close to 1 would signify maturity or stagnation (as most indicators stabilize or decline in maturity). This practical example demonstrates how HELIOS allows the integration of recent empirical metrics into a single index, facilitating comparisons between technologies and assessments of their future trajectory.

# 15 Mathematical Foundations for HELIOS Model Improvement

To overcome the limitations identified in the original HELIOS model, the integration of advanced mathematical methodologies is proposed to provide the framework with greater accuracy, dynamism and predictive capacity.

## 15.1 Non-Linear Normalization of Indicators

The linear normalization currently employed in HELIOS does not adequately capture the non-linear evolution of technology maturity indicators. The progression of a technology, from conception to maturity, is rarely linear; rather, it follows growth patterns that accelerate and then decelerate, forming an S-curve [2, 3]. Therefore, the adoption of sigmoid functions is proposed for a more.

Linear normalization assumes that each unit increase in a metric, such as investment, has the same weight on maturity, regardless of whether the technology is at an early or mature stage [1]. By using sigmoid functions, the HELIOS model can reflect that a small increase in investment or publications in the early stages, where the growth curve is flatter, has a different impact on maturity than the same increase during the exponential growth phase, where the curve is steeper, or near saturation, where the curve flattens again [2, 4, 5, 6]. This provides a more nuanced and empirically justified representation of the progress of technological maturity.

Sigmoid functions, also known as S-curves, are mathematical models that transform any real input value into a value within a bounded range, typically between 0 and 1 [4, 7]. Their "S" shape is ideal for modeling growth processes that start slowly, accelerate rapidly, and then decelerate toward saturation [2, 3, 4, 5, 6, 6, 8]. Two sigmoid functions relevant to this application are:

Standard Logistic Function: Defined as:

$$\sigma(x) = \frac{1}{1 + e^{-x}}$$

For the normalization of an indicator $X_j$:

$$x_j = \frac{L_j}{1 + \exp(-k_j(X_j - X_{0j}))}$$

where L_j is the upper limit (asymptote, typically 1), k_j is the growth rate or slope of the curve for indicator j, and X_0j is the value of indicator j at the inflection point [4, 8].

Gompertz function: Expressed as:

$$f(x) = a \cdot \exp(-b \cdot \exp(-c \cdot x))$$

For the normalization of an indicator $X_j$:

$$x_j = a_j \cdot \exp(-b_j \cdot \exp(-c_j \cdot X_j))$$

where a_j is the upper asymptote (maximum maturity, typically 1), b_j is the shift parameter along the x-axis, and c_j is the growth rate [5, 9]. Unlike the logistic, the Gompertz function is an asymmetric sigmoid, characterized by slower growth at the beginning and end of the period, and a more gradual approach to the upper asymptote [5].

This asymmetry makes it particularly useful for modeling phenomena where initial growth is very slow before a rapid takeoff, followed by prolonged saturation, as is common in the adoption of new technologies or population growth [5, 10].

The specific parameters of these sigmoid functions (L,k,X_0 or a,b,c) will be estimated empirically from historical data for each indicator using nonlinear regression techniques [9].

The choice between logistic (symmetric) and Gompertz (asymmetric) function allows to adapt the normalization to the specific characteristics of each indicator [5, 9]. For example, the adoption of technologies often follows a Gompertz curve, with a very slow initial phase before a rapid take-off [5, 10], whereas the growth of scientific publications might be more symmetric.

This flexibility in the choice of sigmoid function for each variable (I,P,Pt,A,R) allows for a more accurate calibration and a more faithful representation of their individual contribution to maturity, improving the overall robustness of the model

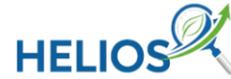

**Table 1: Comparación de Estrategias de Normalización de Indicadores para HELIOS**

| Indicator | Actual Linear Normalization (Formula/Description) Proposed Sigmoid Normalization | Actual Linear Normalization (Formula/Description) Proposed Sigmoid Normalization | Actual Linear Normalization (Formula/Description) Proposed Sigmoid Normalization | Actual Linear Normalization (Formula/Description) Proposed Sigmoid Normalization | Actual Linear Normalization (Formula/Description) Proposed Sigmoid Normalization | Actual Linear Normalization (Formula/Description) Proposed Sigmoid Normalization |
|---|---|---|---|---|---|---|
| **Investment** | $I\_textnorm = I/I\_textmax$ (Linear, sections) [1] | $I\_textnorm = L/(1 + exp(−k(I − I\_0)))$ (Logistics /Gompertz) | Captures initial exponential growth and investment saturation [2, 11]. | \$100M (de \$1B máx) | 0.1 | 0.05 (if early stage) |
| **Publications** | $P\_textnorm = P/P\_textmax$ (Linear, sections) [1] | $P\_textnorm = L/(1 + exp(−k(P − P\_0)))$ (Logistics/Gompertz) | Reflects exponential growth of the literature in rapid development phases [1, 12]. | 50 items (from 1000 max) | 0.05 | 0.15 (if accelerated phase) |
| **Patents** | $Pt\_textnorm = Pt/Pt\_textmax$ (Linear, sections) [1] | $Pt\_textnorm = L/(1 + exp(−k(Pt − Pt\_0)))$ (Logistics /Gompertz) | Models gradual emergence and subsequent acceleration of patents [1, 13, 14]. | 100 patents (from 500 max) | 0.2 | 0.3 (si es fase de crecimiento) |
| **Adoption** | Fixed scale (0-1%)~0, (1-10%)~0.1-0.3, etc. [1] | $A\_textnorm = Mfrac1 − exp(−(p + q)t)1 + (q/p)exp(−(p + q)t)$ (Bass/Gompertz) | Reflects asymmetric adoption pattern (S-curve) [1, 5, 10]. | 5% market | 0.2-0.5 | 0.1 (if the initial phase is very slow) |
| **Regulation** | Direct (0=nulo, 0.5= development, 1=complet) [1] | $R\_textnorm = L/(1 + exp(−k(R − R\_0)))$ (Logistics, with qualitative interpretation) | Captures the non-linear progression of regulatory maturity [1]. | Specific guides (0.5-0.8) | 0.5-0.8 | 0.6 (if intermediate phase) |

## 15.2 Growth and Diffusion Models (S-Curves)

To provide HELIOS with a robust predictive capability, it is essential to model the growth trajectories of its key indicators using S-functions, which are inherent to technological evolution [2, 3, 15]. These curves are characterized by an initial slow phase (genesis), a phase of rapid growth (acceleration) and a final phase of deceleration and saturation (maturity) [2, 3, 8].

S-curves not only describe the cumulative growth, but their derivatives provide crucial information about the speed and acceleration of technological development [11, 13, 16]. The inflection point of the S-curve, where the first derivative is maximum, represents the "tipping point" of the technology, indicating the phase of highest growth [1, 8, 17]. The slope of the S-curve can also reveal the "productivity of R&D investments" [11]. By quantifying these points and rates of change, HELIOS transforms from a descriptive to a predictive tool with clear time milestones, which is invaluable for making strategic investment and development decisions [1, 8, 17].

Modeling each key indicator (Publications, Patents, Adoption) with its own S-curve allows a granular and temporal view of technological maturity [1]. This is superior to a single composite index, as it reveals imbalances. For example, a technology could show a peak in publications and patents, but still incipient adoption. This situation would indicate that the technology is in a "valley of disillusionment" or "early development" phase despite scientific progress [1]. This multifaceted diagnostic capability provides a deeper understanding of the technology's trajectory and helps to identify areas that require further attention or investment to bridge "the chasm" toward mass adoption [1, 18].

For the **Adoption (A), the Bass Diffusion Model is a** classic and widely used approach to forecast the adoption of new products and innovations [1, 10, 19, 20]. This model distinguishes between two types of adopters: innovators (p) and imitators (q) [19, 20]. The adoption rate is formulated as:

For Adoption (A), Bass Diffusion Model:

$$\frac{dN(t)}{dt} = \left[ p + q\frac{N(t)}{M} \right][M - N(t)]$$

where N(t) is the cumulative number of adopters at time t, and M is the total potential market size or saturation level [10, 19, 20]. Cumulative adoption is expressed as:

Cumulative adoption is expressed as:

$$N(t) = M\frac{1 - \exp(-(p+q)t)}{1 + (q/p)\exp(-(p+q)t)}$$

The inflection point, which indicates the point of maximum adoption rate, is calculated as $t\_textinflexion =$

$fracln(q/p)p + q$ [19]. The parameters p, q and M are estimated from historical adoption data by nonlinear regression [10, 19, 20].

For Scientific Publications (P) and Patents (Pt), whose growth also follows an S-curve [1, 12, 13, 14, 21], Logistic or Gompertz Models can be used. The formulation of the Logistic Model for Y(t) (cumulative number of publications or patents) is:

For Publications (P) and Patents (Pt), Logistical Model:

$$Y(t) = \frac{K}{1 + \exp(-r(t - t_0))}$$

where K is the carrying capacity or saturation level, r is the growth rate and t_0 is the time at the inflection point [8, 12, 21]. For the Gompertz model, the formulation is:

Gompertz model:

$$Y(t) = a \cdot \exp(-b \cdot \exp(-c \cdot t))$$

with a as the upper asymptote, b as the shift parameter and c as the growth rate [5, 9]. The parameters of these models are estimated using nonlinear regression on historical data [12, 14, 21].

Once the parameters of the S-models have been estimated from historical data, these functions can be extrapolated to forecast future values of I, P, Pt and A. This allows the generation of quantitative projections for each indicator, which in turn will feed a HELIOS index projected over time.

**Table 2: Key Parameters and Forecasts of S-Curve Models for HELIOS Indicators**

| Indicator | Chosen S-Curve Model | Estimated Key Parameters (Example) | Range of Historical Data Used | Predicted Tipping Point Year | Predicted Saturation Level (Value and Estimated Year) | 5-Year Forecast Value |
|---|---|---|---|---|---|---|
| **Publications** | Logistics | K=1500, r=0.2, t0=2028 | 2000-2024 | 2028 | 1500 articles (2040) | 1000 articles |
| **Patents** | Gompertz | a=700, b=15, c=0.1 | 2005-2024 | 2030 | 700 families (2045) | 450 families |
| **Adoption** | Bass | M=100%, p=0.01, q=0.3 | 2010-2024 | 2032 | 100% market (2050) | 25% market |
| **Investment** | Logistics | K=$50B, r=0.15, t0=2027 | 2015-2024 | 2027 | $50 one thousand million (2042) | $35 one thousand million |

**Figure 1: Publication, Patent and Adoption Forecast Curves**

*Conceptual Description: This figure would show three overlapping or adjacent line graphs, each depicting the predicted trajectory of an indicator (Publications, Patents, Adoption) over time, using their respective S-curve models..*

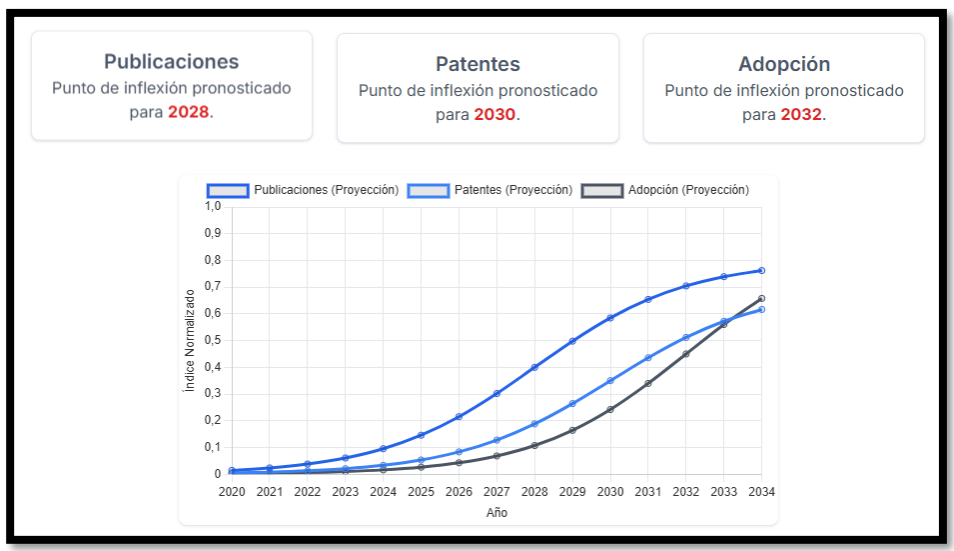

*Illustration 1 Growth Curve Forecasting - Using diffusion models such as Bass and logistic functions, HELIOS projects the future trajectory of each indicator.*

### 4.3 Dynamic and Data Driven Weighting

The assignment of static weights in HELIOS is a simplification that does not reflect the changing importance of indicators over the life cycle of a technology. The fixed weights assigned in the original HELIOS model do not adjust to the evolution of technology [1]. The relevance of investment or publications is very high in the early stages, while adoption and regulation gain weight as the technology matures and diffuses [1, 22]. A dynamic, data-driven weighting scheme is therefore proposed.

The adoption of dynamic and data-driven weights represents a qualitative leap for HELIOS. Instead of a static evaluation, the model becomes sensitive to the context of the technology [23, 24, 25]. This means that the importance of, for example, a new patent, will be evaluated differently if the technology is in its research phase, where a patent is a huge milestone, than if it is in a mature phase, where patents can be incremental. This adaptive approach not only improves the accuracy of the maturity assessment, but also enables a deeper understanding of the key drivers at each stage of the technology lifecycle, which is critical for strategic decision making.

Data-driven weighting methods such as Principal Component Analysis (PCA) or Factor Analysis (FA) can be used to determine the weights. PCA is a dimensionality reduction technique that identifies underlying patterns in sets of correlated variables and can derive objective weights based on the proportion of variance explained by the principal components [26, 27, 28]. The "loadings" of each factor on the first principal component (PC1) can represent the weights, computed as:

for PCA:

$$w_j = \frac{|loading_{j,PC1}|}{\sum_{k=1}^{n} |loading_{k,PC1}|}$$

where n is the number of indicators [27]. FA, similar to PCA, focuses on identifying latent variables that influence the observed variables and is also used to derive weights in composite indices [28, 29]. The choice between PCA and Factor Analysis for the derivation of weights is not merely technical, but has theoretical implications [28]. PCA seeks to maximize explained variance and reduce dimensionality. If "technological maturity" is conceptualized as a latent construct driving the indicators, FA could provide a more robust theoretical validation for the weights. If the primary objective is data reduction and identification of the combinations of indicators that best summarize the information, PCA might be more straightforward. An expert analysis would consider this distinction to ensure that the weighting method chosen aligns with the theoretical conceptualization of the HELIOS model.

In addition, adaptive (dynamic) weighting schemes will be implemented, where the weights are adjusted based on the phase of the technology life cycle [23, 24, 25, 30]. This allows the model to more accurately reflect the changing importance of each indicator as the technology matures [1]. Life cycle phases (e.g., "Emergence/Concept", "Initial Development", "Early Adoption", "Growth/Expansion", "Maturity/Saturation") will be defined based on HELIOS index thresholds or individual indicators [1, 31, 32].

For each phase, a different set of weights (w_textphase,i) will be applied. For example, in the emergence phase, a higher weight would be assigned to investment and publications, and a lower weight to adoption and regulation. In the growth phase, adoption and patents would gain importance, while regulation would increase its weight..

At maturity, adoption and regulation would be more deterministic, and investment, publications and patents could stabilize or decrease [1]. These phase weights can be determined by refined expert judgment, where experts assign specific weights for each life cycle phase rather than a single static set [24, 25, 29]. If sufficient historical data is available for various technologies throughout their life cycles, machine learning algorithms or optimization techniques can be employed to learn the optimal weights for each phase, which could involve predictive models that adjust the weights in real time [33, 34].

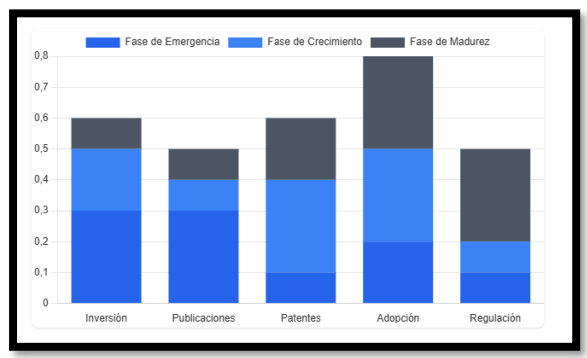

*Illustration 4 Dynamic and Adaptive Weighting - The importance of each indicator changes throughout the life cycle of a technology. HELIOS dynamically adjusts the weights to reflect which factors are most critical in each phase.*

**Table 3: Proposed Dynamic Weighting Scheme for HELIOS by Life Cycle Phase**

| Technology Life Cycle Phase | Proposed Investment Weights (I) | Proposed Weights for Publications (P) | Proposed Patent Weights (Pt) | Proposed Weights for Adoption (A) | Proposed Weights for Regulation (R) | Justification Rational |
|---|---|---|---|---|---|---|
| **Emergence/ Concept [1]** | 0.35 | 0.30 | 0.10 | 0.05 | 0.05 | Investment and publications are critical for initial viability and knowledge exploration. [1]. |
| **Initial Development [1]** | 0.30 | 0.30 | 0.15 | 0.10 | 0.05 | Emphasis on research and early development, with increasing patenting activity. [1]. |
| **Early Adoption [1]** | 0.20 | 0.20 | 0.25 | 0.30 | 0.05 | Adoption is becoming a key factor, along with intellectual property protection. [1]. |
| **Growth/Expansion [1]** | 0.15 | 0.15 | 0.20 | 0.40 | 0.10 | Mass adoption drives growth; regulation begins to be relevant to scalability [1]. |
| **Maturity/Saturation [1]** | 0.10 | 0.10 | 0.15 | 0.35 | 0.30 | Market stabilizes, regulation and standards are key to consolidation [1]. |

*Note: Weights are conceptual and should be calibrated with empirical data and/or refined expert judgment.*

**Figure 2: Dynamic Weighting of Indicators by Life Cycle Stage**

*Conceptual Description: This figure would illustrate how the weights of each indicator (Investment, Publications, Patents, Adoption, Regulation) change through the different phases of the technology life cycle (Emergence, Growth, Maturity). It could be a stacked bar chart or radar chart for each phase, showing the distribution of weights.*

# 16 Nonlinear Aggregation Functions and Synergy Modeling

Simple linear aggregation in HELIOS, using weighted average ($HELIOS = \sum w_i x_i$), assumes a perfect compensatory relationship between indicators [1, 35, 36]. This implies that a low performance in one indicator can be completely compensated by a high performance in another. However, in technological maturity, this assumption is not always valid; for example, the absence of an essential regulatory framework cannot be "compensated" by massive investment if the technology requires a legal environment for widespread adoption. Moreover, linear aggregation does not model synergies, where the combined effect of several indicators is greater than the sum of their parts [36, 37]. For more sophisticated modeling, nonlinear aggregation operators will be explored.Technological maturity is not simply a linear sum of its components; there are complex interdependencies, threshold effects, and synergy or redundancy phenomena [35, 36]. By moving from a linear weighted average to nonlinear aggregation operators such as the Choquet Integral or OWA operators, HELIOS can more accurately reflect these complexities [36, 38, 39]. This allows the model to distinguish, for example, between a technology that makes balanced progress on all fronts, which might be valued more positively by the Choquet Integral, and one that has exceptional performance in one area but critical deficiencies in another. This ability to model non-compensatory relationships is fundamental to a truly expert and nuanced maturity assessment.

Among the advanced nonlinear aggregation operators, the following stand out:

**Choquet Integral:** This is a non-additive aggregation operator that allows modeling interactions (synergies or redundancies) between criteria [36, 38]. It uses "fuzzy measures" or "capacities" that assign importance not only to individual criteria, but also to their combinations [38]. A key advantage is that it can penalize unbalanced achievements, e.g., high investment but low adoption, and capture synergies. For example, high investment and an emerging regulatory framework could have a super-additive effect on perceived maturity, as regulation reduces investment risk [36]. The conceptual formula for a finite set of indicators is.

$$C_\mu(x_1, \dots, x_n) = \sum_{i=1}^{n} x_{(i)} \left[ \mu(A_{(i)}) - \mu(A_{(i+1)}) \right]$$

where $x_{(i)}$ are the normalized values ordered in a non-decreasing order, $A_{(i)}$ is the set of the i largest indicators, and mu is a fuzzy measure that assigns a value to each subset of indicators [38].

**Ordered Weighted Averaging Operators (OWA - Ordered Weighted Averaging):** This is a parameterized class of aggregation operators that allows modeling linguistic aggregation instructions, such as "at least half of the criteria are met" [39, 40]. OWA operators provide a spectrum of aggregation behaviors, from minimum to maximum, including arithmetic mean and median, by adjusting a set of weights [39]. They allow an "attitude" (optimistic or pessimistic) to be incorporated into the aggregation [39]. The formula is:

$$OWA(a_1, \dots, a_n) = \sum_{j=1}^{n} w_j \, b_j$$

where $b_j$ is the j-th largest value of the arguments $a_i$ (the normalized values of the indicators), and $w_j$ are the OWA weights that sum to 1 [39].

**Fuzzy Logic Based Aggregation:** This approach allows handling imprecise or qualitative information, such as that of the regulation indicator [41, 42, 43]. It uses inference rules (e.g., IF (Investment is HIGH AND Adoption is LOW) THEN Maturity is EARLY GROWTH) to combine indicator values [42]. Its advantages include robustness to uncertain and imprecise data, the ability to model complex and nonlinear relationships without the need for explicit mathematical functions, and its interpretability [42, 43]. The application of fuzzy logic or fuzzy measures within the Choquet Integral could transform the qualitative input of "Regulation" into a more nuanced and mathematically tractable representation [41, 42, 43]. Instead of a simple 0, 0.5 or 1, degrees of membership could be assigned to fuzzy sets such as "Incipient Regulation", "Developing Regulation" or "Consolidated Regulation". This not only reduces subjectivity, but also allows the information on regulation to interact in a more sophisticated way with the other indicators in the aggregation process, improving the mathematical consistency of the model.

**Table 4: Comparison of Aggregation Methods for the HELIOS Index**

| Aggregation Method | Mathematical Formula (Conceptual) | Key Properties/ Advantages | Disadvantages/ Complexity | Relevance to HELIOS |
|---|---|---|---|---|
| **Weighted Arithmetic Mean** | $textHELIOS = sum w_i x_i$ [1] | Simple, easy to calculate and interpret. | Asume compensación perfecta, no modela sinergias/redundancias [35, 36]. | Current baseline, useful for quick comparisons if interactions are minimal. |
| **Choquet Integral** | $C\_mu(x\_1, dots, x\_n) sum x\_(i)[mu(A\_(i)) - mu(A\_(i+1))]$ [38] | Handles synergies and redundancies, penalizes imbalances, allows for diffuse measures [36, 38]. | Requires estimation of fuzzy measurements (capabilities), more complex [38]. | Ideal for capturing complex interdependencies between I, P, Pt, A, R and assessing balanced maturity. |
| **OWA Operators** | $textOWA(a\_1, dots, a sum w\_j b\_j$ [39] | Allows to model attitudes (optimistic/pessimistic), flexible between min and max. [39]. | Requires sorting of values, selection of weights OWA [39]. | Useful for incorporating preferences in aggregation or reflecting different evaluation strategies. |
| **Fuzzy Logic** | Based on "IF-THEN" rules and membership functions [42, 43] | Robust for uncertain/qualitative data, interpretable, models nonlinear relationships implicitly [42, 43]. | Requires definition of rules and fuzzy sets, can be data intensive [42]. | Improves quantification of the regulation and allows for more intuitive aggregation of qualitative information. |

# 17 Probabilistic Forecasting and Uncertainty Quantification

For the HELIOS model to be truly predictive and useful for strategic decision making, it must incorporate uncertainty quantification, providing probabilistic forecasts rather than point estimates. Single-point forecasts do not reflect the inherent uncertainty in predictions, especially in complex fields such as technology forecasting [44, 45]. Uncertainty Quantification (UQ) is essential to understand the range of possible outcomes, confidence in predictions, and to identify which uncertainties are most critical [44, 45].

The integration of probabilistic forecasting and uncertainty quantification elevates the HELIOS model from an assessment tool to a high-level decision support system [44, 45]. Decision makers will not only know what the projected maturity level is, but also how confident they can be in that prediction and what the range of possible future trajectories is. This approach can be particularly useful for risk management in contexts of high uncertainty, resource allocation and strategy formulation in environments of high technological uncertainty, enabling more informed and resilient decisions.

Methods for Uncertainty Quantification in HELIOS include.:

**Monte Carlo Simulations (MCS):** MCS is a widely used sampling-based approach for uncertainty quantification and propagation [44, 46, 47, 48, 49]. It allows a distribution of possible future HELIOS index values to be generated, rather than a single value [44, 46]. The process involves:

**Definition of Uncertainty Distributions:** Probability distributions will be defined for model input parameters (e.g., uncertainty in p,q,M of the Bass model; in K,r,t_0 of the logistic models; or in the normalized values of indicators based on historical variance or expert elicitation) [46, 47].

**Repeated Sampling:** Thousands or millions of simulations will be performed, drawing random values from these uncertainty distributions for each parameter [46, 47].

**HELIOS Index calculation:** For each simulation, the full HELIOS index will be calculated using nonlinear normalization, dynamic weights and nonlinear aggregation.

**Distribution of Results**: The resulting collection of HELIOS values will form a probability distribution that represents the probabilistic forecast of the index [46]. The advantages of MCS include its ease of interpretation, robustness, and ability to handle non-Gaussian distributions and correlated inputs [44, 46, 47]. The results allow the generation of prediction intervals (e.g., a 95% confidence interval for the HELIOS index) and estimation of the probability of reaching certain maturity thresholds [45, 48, 50].

**Bayesian methods:** These methods provide a rigorous framework for parameter estimation and UQ, incorporating prior knowledge (priors) and updating beliefs with new data [51, 52]. They can directly estimate posterior distributions of model parameters (e.g., p,q from the Bass model), which are then used to generate probabilistic forecasts. Their advantage lies in providing a more complete UQ, especially useful when data are limited, by allowing the incorporation of expert information or analogues through prior distributions [47, 51].

The **Sensitivity Analysis** is an extension of UQ that identifies which input uncertainties (e.g., variability in a specific weight or uncertainty in an S-curve parameter) contribute most to the uncertainty in the HELIOS index output [48, 49]. This helps to focus model improvement efforts and data collection on the most significant sources of uncertainty [44, 49]. It is often performed simultaneously with UQ using Monte Carlo methods [49].

The forecasting literature suggests that combining forecasts from different methods or models, such as different S-curve models for the same indicator or different weighting approaches, can significantly improve accuracy and reduce errors [53, 54]. This "ensemble forecasting" or "combined forecasting" strategy [53] can be applied to HELIOS: instead of choosing a single model for each component (normalization, S-curves, weighting, aggregation), multiple variations could be run and their probabilistic results combined. This adds an additional layer of sophistication and robustness, mitigating the risk of relying on a single set of assumptions or the calibration of a specific model.

**Figure 3: Probabilistic Forecasting of the HELIOS Index over Time**

*Conceptual Description:* The X-axis represents time in years, while the Y-axis represents the HELIOS Index score, which varies from 0 to 1. The solid center line shows the median HELIOS Index forecast, indicating the most likely trajectory of technological maturity. Around this center line, shaded areas are displayed that represent prediction intervals, such as 50% and 95% confidence. These shaded areas illustrate the range of possible outcomes for the HELIOS index at each future point in time, reflecting the inherent uncertainty in the predictions. The width of these shaded areas can vary, widening as the forecast extends further into the future, indicating an increase in uncertainty. This visualization is crucial for communicating uncertainty effectively to decision makers, as it provides a more complete understanding of the risks and opportunities associated with a technology's trajectory.

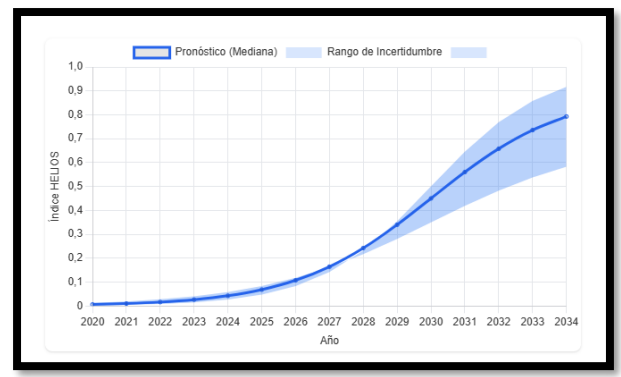

*Illustration 5 Advanced Aggregation and Uncertainty Quantification The model uses the Choquet Integral to model synergies and penalize imbalances. Through Monte Carlo simulations, HELIOS provides a range of possible futures, allowing for a better understanding of the future.*

## 18 Proposal for an Improved HELIOS Model: Integral Mathematical Formulation

The improved HELIOS model integrates the proposed advanced mathematical formulations to provide more accurate assessment, forecasting and quantification of uncertainty of technological maturity. The HELIOS index (H) at time t is defined as a nonlinear aggregation function of dynamically normalized indicators, with adaptive weights:

$$H(t) = \textbf{Aggregation\_Function}(x_I(t), x_P(t), x_{Pt}(t), x_A(t), x_R(t) | W_{\textbf{dynamic}}(t))$$

Where Aggregation_Function can be the Choquet Integral, an OWA operator or an aggregation function based on fuzzy logic, selected based on the desired interaction properties between the indicators. W_textdynamic(t) represents the set of adaptive weights for the maturity stage of the technology at time t.

The complete algorithm for the calculation, prediction and quantification of the uncertainty of the improved HELIOS index is structured in the following steps:

7. **Non-linear Normalization of Indicators**:
   For each indicator X_j (Investment, Publications, Patents, Adoption, Regulation), a sigmoid function f_j (Logistic or Gompertz), calibrated with historical data, is applied. For example, for a logistic normalization:

   $$x_j(t) = \frac{L_j}{1 + \exp(-k_j(X_j(t) - X_{0j}))}$$

   The parameters L_j,k_j,X_0j are pre-estimated for each indicator j from its historical data.

   **Growth Modeling and Indicator Forecasting (S-Curves):**
   For the indicators of Publications (P), Patents (Pt) and Adoption (A), S-curve models (Logistic, Gompertz or Bass) are fitted to their historical data. These models allow predicting the future values of X_j(t) for ttextcurrent_time.

   **Adoption (A): The Bass Model is used.:**

   $$A(t) = M_A \frac{1 - \exp(-(p_A + q_A)t)}{1 + (q_A/p_A)\exp(-(p_A + q_A)t)}$$

   **Publications (P): A Logistic Model can be employed:**

   $$P(t) = \frac{K_P}{1 + \exp(-r_P(t - t_{0P}))}$$

   **Patents (Pt): A Logistic or Gompertz Model is also adequate.:**

   $$P(t) = \frac{K_{Pt}}{1 + \exp(-r_{Pt}(t - t_{0Pt}))}$$

   The parameters $(M\_A, p\_A, q\_A, K\_P, r\_P, t\_0P, K\_Pt, r\_Pt, t\_0Pt)$ are calibrated with historical data.

8. **Determination of Dynamic Weights**:
   The current stage of the technology (e.g., "Emergence", "Growth", "Maturity") is identified based on the current HELIOS index value or key indicator thresholds. The set of weights predefined or derived by PCA/FA for the derived by PCA/FA for that phase is applied.:

   $$Wdynamic(t) = \{wI, fase, wP, fase, wPt, fase, wA, fase, wR, fase\}$$

9. **3. Non Linear Aggregation**:
   Normalized values $x\_j(t)$ and dynamic weights $W\_textdynamic(t)$ are combined using a nonlinear aggregation function. For example, if the Choquet Integral is chosen, the definition of a fuzzy mu measure that captures the interactions between the indicators is required.

10. **Probabilistic Forecasting and Uncertainty Quantification:**
    To generate probabilistic forecasts and quantify uncertainty, Monte Carlo simulations are used..
    - Probability distributions are defined for the parameters of the curve models in S.($e.g., p\_A, q\_A, M\_A, etc.$), standardization parameters ($L\_j, k\_j, X\_0j$), and, if applicable, fuzzy measures or OWA weights.
    - N iterations (e.g., 100,000) of the simulation are performed. In each iteration k:
      - The parameters of the defined distributions are randomly sampled..
      - Indicator values are predicted. $X\_j(t)\_k$ for the desired time horizon.
      - Normalized $x\_j(t)\_k$.
      - The phase of the technology is determined and dynamic weights are applied. $W\_textdynamic(t)\_k$.
      - The index is calculated $H(t)\_k$ using the nonlinear aggregation function.
    - The N values resulting from. $H(t)\_k$ for each t form a probability distribution, from which median, confidence intervals (e.g., 95%) and other uncertainty metrics can be derived.

11. **Sensitivity Analysis:**

    Subsequently, a sensitivity analysis is performed on the Monte Carlo results to identify which of the uncertainties in the input parameters have the greatest impact on the variability of the predicted HELIOS index.

# 19 Conclusions and recommendations

The evolution of the HELIOS model, from its original formulation to the proposed advanced framework, represents a relevant methodological evolution with respect to the previous model in the assessment and forecasting of technological maturity. By overcoming the limitations of linearity, stationarity and lack of predictive capability, the model becomes a more accurate tool for strategic decision making. Non-linear normalization using sigmoid functions more accurately reflects the real contribution of each indicator, since the impact of a change in a metric varies according to the phase of the life cycle: an initial breakthrough or an increase in saturation has different meanings than the same change in an exponential growth phase.

The integration of S-growth models for publications, patents and adoption provides quantitative predictive capabilities, allowing the forecasting of inflection points and saturation levels, which are essential for planning R&D and anticipating "take-off points". This granular view also diagnoses imbalances between indicators and critical areas to overcome adoption barriers..

Dynamic weighting, adjusted to each phase of the life cycle, improves the relevance of the assessment and optimizes resource allocation. Finally, nonlinear aggregation functions and uncertainty quantification through Monte Carlo simulations capture interactions and synergies, providing not only predictions, but probabilistic ranges with confidence levels, essential for risk management in high uncertainty environments.

## 19.1 Recommendations:

**Historical Data Collection and Calibration:** It may be beneficial to establish a robust program for the continuous collection of detailed historical data for each indicator (I, P, Pt, A, R) over time and for various technologies. These data are the basis for accurate calibration of sigmoid function parameters, S-growth models, and fuzzy measurements..

**Specific Software Development:** The development of a dedicated software platform implementing the proposed advanced mathematical formulations is recommended. This platform should allow data input, model calibration, running Monte Carlo simulations, visualizing probabilistic forecasts, and performing sensitivity analysis..

**Validation and Continuous Refinement:** The model must undergo a process of continuous validation and refinement. This includes comparison of the model's forecasts with the actual evolution of technologies, periodic revision of dynamic weights and aggregation functions, and incorporation of new data and expert knowledge to improve accuracy and robustness..

**Expert Elicitation for Fuzzy Parameters and Fuzzy Measurements:** To maximize accuracy, especially in the absence of sufficient historical data, structured expert elicitation is recommended to define the uncertainty distributions of the initial parameters of the models and to construct the fuzzy measures in the Choquet Integral, if this is chosen.

**Scenario and Sensitivity Analysis:** Use the ability to quantify uncertainty to perform scenario analysis. This will allow decision makers to explore the impact of different assumptions (e.g., changes in investment, new regulations) on the maturity trajectory of a technology and understand which factors are most critical to its success or failure.

By implementing these improvements, the HELIOS model is positioned as an interesting tool with room for improvement for forecasting and managing technological maturity, providing a quantitative and probabilistic basis for strategic decisions in the dynamic innovation landscape..

## 20 Open model

The enhancement proposals we have outlined represent a roadmap for taking the HELIOS model to its next phase of evolution. Building on the mathematically rigorous foundation already established, these extensions seek to transform the model from a static framework to a robust, dynamic, predictive tool capable of reflecting the complexity inherent in the life cycle of science..

To achieve this goal, we believe that continuous and open collaboration is essential. First, rigorous validation and calibration of the model is a critical step. We propose the use of backtesting and cross-validation techniques to ensure that the model not only works in theory, but also has a demonstrable predictive capability with historical data. In addition, incorporating a Bayesian approach to the calibration of its parameters will allow us not only to make predictions, but also to quantify the uncertainty associated with them, providing a more solid basis for decision-making.

Secondly, the richness of the information lies not only in the numbers, but also in the context. Therefore, we suggest the integration of unstructured data and network analysis. The use of Natural Language Processing (NLP) techniques to analyze texts from publications, patents and news could extract new latent variables, such as sentiment or research direction. At the same time, the analysis of networks, such as co-authorship and citation networks, would reveal the structure and dynamism of the scientific community, identifying key players and the interconnections that drive innovation.

Finally, to overcome the limitations of fixed formulations, we propose the adoption of machine learning models. Instead of relying on a single formula, ensemble models such as Random Forest or Gradient Boosting can capture the complex nonlinear interactions between indicators. Furthermore, for optimal management of science investment portfolios, multi-objective optimization would allow us to go beyond simple scoring, balancing multiple factors such as potential impact, risk and technological diversity.

We are convinced that these extensions, in collaboration with the community, can transform HELIOS into an advanced tool. We therefore invite the academic community, data analytics experts and industry professionals to join us in validating, refining and expanding this model. Together, we can build an evaluation system that not only measures the past, but also provides an improved framework for prospective analysis in science and technology.

## 21 Statement on the use of AI

During the development of this document, an open source language model (LLMO) was used, executed in a local environment, corresponding to the version available in May 2025. This tool was used to extend and refine the basic HELIOS model, generating drafts and exploring alternative formulations. It was not involved in critical analysis or interpretation of results. Tests were also carried out with real data using this same model, with the aim of obtaining experimental results that were also incorporated in the publication of the book by the same author: *Horizontes Disruptivos: Las 10 tecnologías que redefinirán la próxima década* (ISBN 9798294856489). All the content generated was reviewed, edited by the author.

## 22 REFERENCIAS